\documentclass[aps,prb,twocolumn,superscriptaddress,showpacs,preprintnumbers,amsmath,amssymb]{revtex4-2}
\usepackage{graphicx}% Include figure files
\usepackage{bm}% bold math
\usepackage{enumerate}
\usepackage{mathrsfs}
\usepackage{float}
\usepackage{url}
\usepackage{makecell}
\usepackage{natbib}
\usepackage{multirow}
\usepackage[titletoc]{appendix}
\usepackage[final]{changes}
\definechangesauthor[name={Q.Yao}, color=red]{QY}
\definechangesauthor[name={Askar A. Iliasov}, color=orange]{As}
\usepackage{color}

\usepackage{bookmark}
\usepackage{natbib}
\usepackage{hyperref}
\hypersetup{nolinks=false,
urlcolor=cyan, % url
citecolor=violet, % citation
anchorcolor=yellow,
linkcolor=blue, % table of contents, inner color
citecolor=red,
colorlinks=true}
\usepackage{ragged2e}
\usepackage{multirow}
\usepackage{eucal}
\begin{document}
%\bibliographystyle{apsrev4-2} % Choose Phys. Rev. style for bibliography, Rev.4
%\preprint{APS/123-MW}
%\title{Disorder-free many-body localization in plane fractal}
\title{Level-spectra Statistics in Planar Fractal Tight-Binding Models}
%\title{Level Clustering in Planar Fractal Tight-Binding Models}
%\title{Level Statistics in Sierpinski Fractal: Intrinsic and Disorder-induced traits}
%\title{Many-body Critical Phase in Hierarchical Complex System}
%\title{Many-body Critical Phase in Hierarchical Sierpinski Lattice}

\author{Qi Yao\href{https://orcid.org/0000-0002-0522-2820}{\includegraphics[scale=0.06]{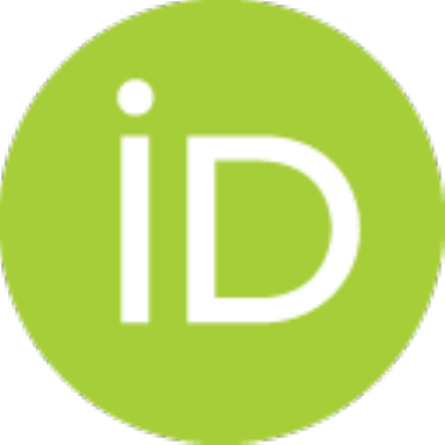}}}
%\thanks{These authors contributed equally to this work.}
\affiliation{Key Laboratory of Artificial Micro- and Nano-structures of Ministry of Education and School of Physics and Technology, Wuhan University, Wuhan, Hubei 430072, China}
\affiliation{Institute for Molecules and Materials, Radboud University, Heijendaalseweg 135, 6525 AJ Nijmegen, The Netherlands}

\author{Xiao-Tian Yang\href{https://orcid.org/0000-0002-5780-6166}{\includegraphics[scale=0.06]{ORCIDiD.pdf}}}
%\thanks{These authors contributed equally to this work.}
\affiliation{Key Laboratory of Artificial Micro- and Nano-structures of Ministry of Education and School of Physics and Technology, Wuhan University, Wuhan, Hubei 430072, China}
\affiliation{Institute for Molecules and Materials, Radboud University, Heijendaalseweg 135, 6525 AJ Nijmegen, The Netherlands}

\author{Askar A. Iliasov\href{https://orcid.org/0000-0003-2409-7292}{\includegraphics[scale=0.06]{ORCIDiD.pdf}}}
\affiliation{Institute for Molecules and Materials, Radboud University, Heijendaalseweg 135, 6525 AJ Nijmegen, The Netherlands}
\author{M. I. Katsnelson\href{https://orcid.org/0000-0001-5165-7553}{\includegraphics[scale=0.06]{ORCIDiD.pdf}}}
\affiliation{Institute for Molecules and Materials, Radboud University, Heijendaalseweg 135, 6525 AJ Nijmegen, The Netherlands}

\author{Shengjun Yuan\href{https://https://orcid.org/0000-0001-6208-1502}{\includegraphics[scale=0.06]{ORCIDiD.pdf}}}
\email[Corresponding author: ]{s.yuan@whu.edu.cn}
\affiliation{Key Laboratory of Artificial Micro- and Nano-structures of Ministry of Education and School of Physics and Technology, Wuhan University, Wuhan, Hubei 430072, China}
\affiliation{Wuhan Institute of Quantum Technology, Wuhan 430206, China}
%\affiliation{Institute for Molecules and Materials, Radboud University, Heijendaalseweg 135, 6525 AJ Nijmegen, The Netherlands}

%\date{\today}
\date{June 15, 2022}

\begin{abstract}
   In this communication, we study the level-spectra statistics when a noninteracting electron gas is confined in \textit{Sierpi\'{n}ski Carpet} (\textit{SC}) lattices. These \textit{SC} lattices are constructed under two representative patterns of the $self$ and $gene$ patterns, and classified into two subclass lattices by the area-perimeter scaling law. By the singularly continuous spectra and critical traits using two level-statistic tools\iffalse the nearest spacing distribution and alternative gap-ratio distribution\fi, we ascertain that both obey the critical phase due to broken translation symmetry and the long-range order of scaling symmetry. The Wigner-like conjecture is confirmed numerically since both belong to the Gaussian orthogonal ensemble. An analogy was observed in a quasiperiodic lattice~\cite{Zhong1998Level}. In addition, this critical phase isolates the crucial behavior near the metal-insulator transition edge in Anderson model. The lattice topology of the self-similarity feature can induce level clustering behavior.
\end{abstract}

%\pacs{03.75.Mn, 03.67.Pp, 76.60.Lz}
\maketitle
%%%%%%%%%%%%%%%%%%%%%%%%%%%%%%%%%%%%%%%%%%%%%%%%%%%%%%%%%%%%%%%%%%%%%%%%%%%%%%%
\section{Introduction}
Fractals, as widely known, are flawless from two perspectives. The first is the irregular geometry of natural objects such as clouds and mountains~\cite{Mandelbrot1982fractal,schroeder2009fractals,falconer2004fractal,nakayama2003fractal,Mandelbrot1977fractals}. These shapes can be either determinedly or statistically invariant on an arbitrary scale. Some representative cases are \textit{Cantor set}~\cite{Mandelbrot1982fractal}, \textit{Sierpi\'{n}ski lattice}~\cite{Sierpinski1915Sur,falconer2004fractal}, \textit{Vicsek lattice}~\cite{Vicsek1983Fractal,You1993Exact} and \textit{Koch curve}~\cite{von1906methode}, even boundary modes of lattices~\cite{Even1999Localizations,Wong1992Correlation,Sapoval1993Vibrations,Kopelman1997Spectroscopic}, and random regular graphs~\cite{Metz2021Localization}. The second is the self-similarity behaviors of quantum counterparts, which exist in the Hofstadter butterfly spectra~\cite{Matsuki2021Fractal} and electronic properties in terms of level distribution~\cite{You1990Dynamical}, density of state (DoS)~\cite{Pedersen2020Graphene}, and wavefunction~\cite{Wojcik2000Time}.

Some researchers advocated fractals and quasicrystals as the intermediate between crystalline and amorphous systems~\cite{Rammal1983Nature,Schwalm1989Electronic,Schwalm1993Explicit}. Both keep long-range orders, although they lose translational symmetry. Primarily, 1D Fibonacci chain and Penrose tiling~\cite{Penrose1874The,Martin1977math} in 2D space are suggested theoretically as quasiperiodic lattices. Quasicrystal Al-Mn alloy was first observed~\cite{Shechtman1984Metallic} by diffraction patterns, which implies multifold rotation symmetry~\cite{Levine1984Quasi}.

Recently motivated by engineered fractals~\cite{George2006Nanoassembly,Fan2014Fractal,Hernando2015Mesoscale,Shang2015Assembling,Zhang2016Robust,Kempkes2018Design,Liu2021Sierpi} in experiments and several crucial works~\cite{Gefen1980Critical,Ohta1988Growth}, fractals are attracting growing interest in terms of (i) further classification using many topology characters, including topological dimension~\cite{Mandelbrot1982fractal}, the order of ramification~\cite{Gefen1980Critical,Yu1988Numerical}, connectivity~\cite{Mandelbrot1982fractal,Suzuki1983Phase}, lacunarity~\cite{Gefen1983Geometric,Lin1986suggested,Taguchi1987Comment,Lin1987Classi,Wu1988Comment}, etc.; (ii) material properties such as quantum transport~\cite{Edo2016Transport,Xu2021Quantum}, Hall conductivity~\cite{Askar2020Hall}, optical spectra~\cite{Edo2017Optical}, and plasmon mode~\cite{Tom2018Plasmon}; and (iii) other particles such as spin~\cite{Gefen1980Critical,Gefen1984Phase}, photons~\cite{Akkermans2010Thermodynamics}, and harmonic oscillator~\cite{Sapoval1993Vibrations,Rammal1984Spectrum} upon these lattices.
Here we focus on \textit{Sierpi\'{n}ski Carpet}~\cite{Mandelbrot1977fractals,Gefen1980Critical,Mandelbrot1982fractal,Gefen1984Phase} in two broader families of construction, whose order of ramification is infinite, and it resembles a translation-invariant square lattice in $2D$. It is fabricated perhaps by an optical waveguide array~\cite{Pal2013Engineering,Xia2022Exact} or photonic lattices~\cite{Xu2021Quantum} at medium-scale ($g\in[3,5]$, see $g$ in the context below), or more extreme scenarios such as local-field tailoring~\cite{Yang2020Confined}. \iffalse In this work, we focus on two aspects: (i) How much does it affect the nature of the effective fractal about by local-field modulation? (ii) Under this effect, what happens to\fi

%%%%%%%%%%%%%%%%%%%%%%%%%%%%%%%%%%%%%%%%%%%%%%%%%%%%%%%%%%%%%%%%%%%%%%%%%%%%%%%%%%%%%%%%%%%%%%%%%
%\section{Sierpinski Carpet}
\textit{Sierpi\'{n}ski Carpet.}--
Let us define a \textit{Sierpi\'{n}ski Carpet} lattice with $\textit{SC}(\mathrm{n}, \mathrm{m}, \mathrm{g})$, which can be experimentally carried out by the similar approaches above~\cite{Pal2013Engineering,Xia2022Exact,Xu2021Quantum}. The hierarchy level $g$ is about scale invariance~\cite{Mandelbrot1982fractal,falconer2004fractal}, which is viewed vividly when one scales these structures. Before constructing \textit{SC} lattices, we introduce the $generator(n,m)$ and dilation pattern. The former is the ``seed" lattice, which is constructed by the following steps: one first has a square lattice of size $n$ by $n$, and then dips a subsquare-block lattice of $m$ by $m$ from its center~\cite{Alexander2017The} (here supposing that the $generator(n,m)$ is symmetrical; hence $n$ is always larger than $m$, and $n-m$ is an even number. $n$ ($m$) only starts from 3 (1), followed by 4 (2), 5 (1, 3), 6 (2, 4), and so on, see Table~\ref{table:DH}). The dilation pattern selects two representatives of the $self$ and $gene$ pattern (named shortly as $M_{se,ge}$, and both are in matrix form).

\begin{figure}[!htbp]
    %\centering
{\includegraphics[width=8.5cm,height=4.25cm]{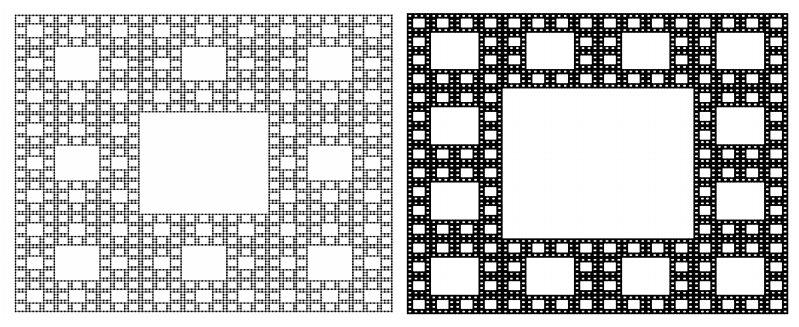}}
\caption{(Color online) Pictorial sketch of pre-fractals in the \textit{SC} family $SC(n,m,g)$ by a paradigmatic $generator(4,2)$ under \textit{self} and \textit{gene} patterns. Corresponding spatial atom-clusters are exhibited in the left and right panel, respectively with hierarchical level $g=3$. The dilation of the $self$ pattern is captured easily by $\mathcal{D}_{se}=\log \mathcal{N}/ \log r$, here we set $\mathcal{N}=8$ and $r=3$ when $g$ increases by 1 for simplification; reversely, the $generator$ pre-determines the $M_{ge}$ (its definition sees the main text). The more cases refer to Table~\ref{table:DH}.}\label{Fig1:Sketch}
\end{figure}

Using the $generator(n,m)$ and the two above patterns, we can dilate two $SC(n,m,g)$ lattices under Eq.~\ref{eq1},
\begin{eqnarray}\label{eq1}
\textit{SC}(\mathrm{n}, \mathrm{m}, \mathrm{g})=M_{se,ge}(g) \otimes \textit{generator}(n, m).
\end{eqnarray}
$g$ is degraded with $M_{se,ge}(g)\equiv M_{se,ge}(g-1)\otimes M_{se,ge}(1)$, and the symbol $\otimes$ is the tensor product of the matrix.
Two things about pattern choices are noteworthy. (i) $M_{se}$ has $r^2$ elements, consisting of $\mathcal{N}$ ones and $r^2-\mathcal{N}$ zeros (for the $SC(n,m,g)$ lattice, its perimeter length is expanded by $r$ times, and its area increases by $\mathcal{N}$ times when $g$ grows by one). Placing these elements flexibly in the matrix $M_{se}$ is possible, however, which easily results in numerous noticeable subclusters in both the spatial lattice and energy spectra (see the $vari$ pattern as a variation of the $self$ pattern in Sup.~\ref{Sup:Density}). The $self$ pattern with the accessible geometry dimension $\mathcal{D}_{se}$ is intensively studied~\cite{Wong1992Correlation,Edo2016Transport}. (ii). Conversely, $M_{ge}$ is given exactly by the graphical mapping of an individual $generator(n,m)$, its size is $n$ by $n$, and all elements consist of $n^2$ ones and $m^2$ zeros. To visualize two pattern classes, a case with the $generator(4, 2)$ in Fig.~\ref{Fig1:Sketch} gives $M_{se}$ (tacitly setting $r=3$ and $\mathcal{N}=8$, and $M_{se}=[1\;1\;1; 1\;0\;1; 1\;1\;1]$.) and $M_{ge}(1)=[1\;1\;1\;1;  1\;0\;0\;1; 1\;0\;0\;1;  1\;1\;1\;1]$. One iterates Eq.~\ref{eq1} ad infinite for $SC(n,m,\infty)$, or instead finitely for pre-fractal $SC(n,m,g)$. These lattices are classified into two scale-invariance classes.

The area-perimeter scaling law~\cite{Mandelbrot1977fractals,feder2013fractals} $Ar\propto p^\mathcal{D}$ discriminates these two $SC(n,m,g)$ classes by dimension index $\mathcal{D}$. Physically, and this quantity defines the filling factor of the $SC(n,m,g)$ lattice in 2D space, sometimes called as \textit{Hausdorff dimension} $\mathcal{D}_H$. The rule is to enumerate the vertex number in $SC(n,m,g)$ lattice, the perimeter length is $p=m r^g$  and the area is $Ar=(n^2-m^2) \mathcal{N}^g$ for the $self$ pattern, and $p=m^{g+1}$  and $Ar=(n^2-m^2)^{g+1}$ for the $gene$ pattern. We note that the exponential factor of $Ar$ is $g$ and $(g+1)$, respectively. This signifies that the geometrical hierarchy is scaled extraordinarily once in the $gene$ pattern (see Fig.~\ref{Fig1:Sketch}, scaling two $SC(4,2,3)$ lattices in the same size, the lattice in the $gene$ pattern has a darker black color than that in the $self$ pattern). To correct this ill definition more physically, we adopt the geometric hierarchy $g^*$ in the following. In addition, in the large $g$ limit, after some algebraic calculations, we derive two asymptotic formulas $\mathcal{D}_{se} = \log (\mathcal{N})/ \log (r)$ and $\mathcal{D}_{ge}=\log(n^{2}-m^{2})/\log(n)$. Interestingly, $\mathcal{D}_{ge}$ only depends on its $generator(n,m)$; instead, $\mathcal{D}_{se}$ is associated with the external self-similarity choice~\cite{Mandelbrot1982fractal}. There is much work concerning $\mathcal{D}$. Van Veen \textit{et al.} reported the box-counting dimension $\mathcal{D}_{box}$ from numerical quantum conductance fluctuation~\cite{Edo2016Transport}, which is coincident with $D_{se}$; again, in a \textit{Sierpi\'{n}ski} lattice sample of CO molecules on a Cu (111) surface~\cite{Kempkes2018Design}. Kempkes \textit{et al.} experientially derived the dimension $\mathcal{D}_{\psi}$ from alternative electronic wavefunctions $\psi$.

Upon these $SC(n,m,g^*)$ lattices, a noninteracting electron gas is described by the model,
\begin{eqnarray}\label{eq2}
H=-\sum_{\langle i,j\rangle}(\mathbf{c}_i^\dagger \mathbf{c}_j+\mathbf{c}_j^\dagger \mathbf{c}_i)+\sum_{i}V \mathbf{c}_i^\dagger \mathbf{c}_i,
\end{eqnarray}
a single Wannier orbit $\mathbf{c}_i^\dagger|0\rangle$ is at site index $i$, and the sign $\langle i,j \rangle$ labels the nearest-neighbor site pair. We set $V=0$, and the energy spectra result purely from the fractal-lattice topology.  We next study how the electronic energy spectra evolve as the $generator(n,m)$ and dimensionality $\mathcal{D}_{se,ge}$ change. $n$, $m$, and $\mathcal{D}_{se,ge}$ refer to Table~\ref{table:DH}.

The spectral correlation between the proximate energy levels gives more insights; here, we use the level-spacing distribution (LSD). As forecasted by random matrix theory (RMT)~\cite{Akemann2015RMT}, three kinds of LSDs are Poisson (Wigner) distributions with the localized (extended) phases, and the intermediate one for the critical phase. Iliasov \textit{et al.} obtained $P(s)\sim s^\alpha$ in the small-$s$ limit in two simplified iterated structures (each square- and triangle-block unit with less connection~\cite{Iliasov2020Linearized}, respectively. These structures facilitate the renormalization group approach when decimating sites). Its numerical diversity~\cite{Askar2019Power} is evident in varying $s$-distance, primarily due to the absence of the unfolding procedure. Essentially, although the model is nonrandom upon these two simple lattices, it still obeys the Wigner-like surmise because of identical Gaussian orthogonal symmetry~\cite{Dyson1962I,Dyson1962II,Dyson1962III}, having the power-law scaling. This assertion should be supported even in these $SC(n,m,g^*)$ fractals. Notably, Zhang \textit{et al.} pioneered this possibility in quasicrystals~\cite{Zhong1998Level}.

This work explores the energy spectra and their LSDs in $SC(n,m,g^*)$ fractals where a single electron roams\iffalse with the pure hopping term of Eq.~\ref{eq2}\fi. We demonstrated that a critical phase (CP) exists in such a system, where all states are critical. Predictably, these states are multifractal in spatial lattice and have rich coherence traits~\cite{Yao2022Wave}. We can consider that the CP is affected by the disorder~\cite{Yao2022New}. Moreover, \textit{level clustering} exists in these fractals, which can be treated as a particular case for level attracting.

This paper is organized as follows. Sec.~\ref{FraLS} covers two primary subsections. First, we review the overall band feature under the action of the patterns ($M_{se}$ and $M_{ge}$), the $generator(n,m)$, and $g^*$. In virtue of the local energy cluster and integrated DoS, the broad energy spectra are diagnosed to be singularly continuous, which hints at a critical phase (note that by the bandwidth trend changes, when all states reside in such a phase, it perhaps cannot make a transition by tuning the $generator(n,m)$ and $g^*$). Second, we analyze the spectral correlation between energy levels by the nearest-level spacing $P(s)$ and the gap-ratio $P(\tilde{r})$. By their numerical traits differing from the extended and localized phases, we assert that electronic states are in a critical phase, and $P(s)$ also differs from that near Anderson transition. Several fittings $P_{fit}(s)$ at various $s$ regions with the lattice size $Ar$ being approximately ${\color{blue}10^4}$ or larger affirm the Wigner-like surmise. A conclusion is reached in Sec.~\ref{Conclu}. Sup.~\ref{Sup:Density} demonstrates the level density leg about three $SC(4,2,2)$ lattices under the $self$, $gene$, and $vari$ patterns. Sup.~\ref{Sup:Unfolding} gives the numerical unfolding procedure for $P(s)$.

\begin{table*}[!htbp]%[!htbp]
\caption{\label{table:DH} Planar $SC(n,m,g^*)$ lattices under peculiar configuration between the $generator(n,m)$ and the $self$ or $gene$ patterns, are assorted by area-perimeter scaling law~\cite{Mandelbrot1977fractals,feder2013fractals} into two subclasses with various Hausdorff dimensions ($\mathcal{D}_{se}$ and $\mathcal{D}_{ge}$, apart in line 2 and line 5). The maximum energy $E_m$ is collected from Fig.~\ref{Fig2:DireEne}(a1), (b1), (c1), and (d1); and the mean gap-ratio $\langle \tilde{r} \rangle$ is derived from the \textbf{\textit{\~{r}}}-value statistic in Fig.~\ref{Fig3:LevelSta}.(b). Here the bracket $(...,...,...)$ adds sequently the quantities $E_m$ and $\langle \tilde{r} \rangle$ with increasing $g^*$ value~\cite{CompuMax} from 2. For clarity, we use an illustration cell (3.2693, 3.3035, 3.3065), the $E_m$ is respectively 3.2369, 3.3035, and 3.3065 in the unit of $t$ for \textit{Sierpi\'{n}ski Carpet} lattice $SC(5,3,2)$, $SC(5,3,3)$ and $SC(5,3,4)$.}
\begin{tabular}{p{0.05\linewidth}p{0.03\linewidth}p{0.2\linewidth}p{0.2\linewidth}p{0.25\linewidth}p{0.15\linewidth}}
        \hline \hline
         \multicolumn{2}{c}{$SC(n,m,g^*$)} & (5, 3) & (4, 2) & (3, 1) & (5, 1) \\ \hline
%        \textit{gene} & 1.7227 &1.7925 & 1.8928 & 1.9746 \\
        \multirow{3}{*}{self}& & 1.8928 &1.8928 & 1.8928 & 1.8928 \\ \cline{3-6}
                             & $E_m$& (3.2693, 3.3035, 3.3065) & (3.2940, 3.3492, 3.4563) &(2.9498, 3.3519, 3.4455, 3.4563) &(3.7857, 3.8263)\\
                             & $\langle \tilde{r} \rangle$ & (0.1437, 0.1353, 0.1433) & (0.1492, 0.1585, 0.1496) & (0.1951, 0.1503, 0.1556, 0.1605) & (0.1830, 0.1912) \\ \hline
        \multirow{3}{*}{gene}&  &1.7227 &1.7925 & 1.8928 & 1.9746 \\ \cline{3-6}
                             & $E_m$&(2.9479, 3.2818)& (2.9250, 3.3050, 3.3484)&(2.9498, 3.3519, 3.4455, 3.4563) & (3.8106, 3.8647) \\
                             & $\langle \tilde{r} \rangle$ & (0.1158, 0.0980) & (0.1419, 0.1231, 0.1150) & (0.1951, 0.1503, 0.1556, 0.1605) & (0.1773, 0.1984) \\
        \hline \hline
        \multicolumn{6}{l}{note: we compute $D_{se}=1.8928$ with setting $\mathcal{N}=8$ and $r=3$.}
        %%\multicolumn{5}{l}{\qquad \; $D_{ge}=\log(n^{2}-m^{2})/\log(n)$} \\
        %\hline \hline
\end{tabular}
%\vspace{-9pt}
\end{table*}

\section{Energy level statistics analysis}\label{FraLS}
\begin{figure}[!htbp]
  \includegraphics[width=8.7cm,height=9cm]{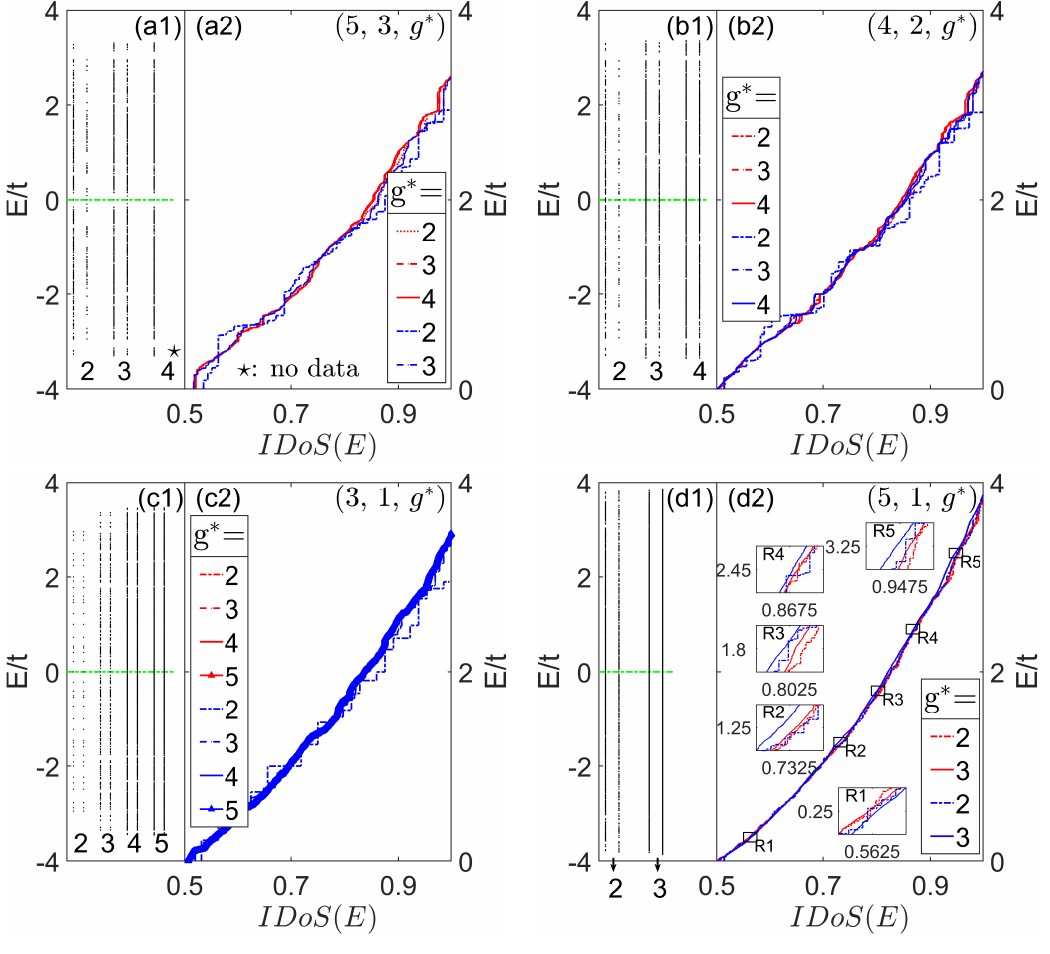}
%  \centering
%   {\includegraphics[width=0.494\columnwidth]{Fig2Level5-3.pdf}}
%   {\includegraphics[width=0.494\columnwidth]{Fig2Level4-2.pdf}}
%   {\includegraphics[width=0.494\columnwidth]{Fig2Level3-1.pdf}}
%   {\includegraphics[width=0.494\columnwidth]{Fig2Level5-1.pdf}}
  \caption{(Color online) Allowed energy spectra and the half-branch integrated DoS of $IDoS$(E) change with a configuration of a $generator(n,m)$ and geometrical hierarchy level $g^*$. In (a1), full band structures are plotted in pairs about the $self$ and the $gene$ pattern, which are related to red and blue color $IDoS(E)$ in (a2) for the $generator(5,3)$, respectively. Similarly, the other three $generator$s $(4,2)$, $(3,1)$, and $(5,1)$ cases in Table~\ref{table:DH} are shown about the band in (b1), (c1), and (d1); and about $IDoS(E)$ in (b2), (c2), and (d2). Each $IDoS(E)$ uses the histogram $\rho(E) =1/Ar \sum_{m}\delta\left(E-E_{m}\right)$ and its integrated region is from -4 to 4. $g^*$ increases to the maximum value~\cite{CompuMax}. The green dot line $E=0$ is added to exhibit level-distribution symmetry. In (d2), five insets ($R1$-$R5$) zoom out $IDoS(E)$ at different positions.}\label{Fig2:DireEne}
\end{figure}

\emph{Level density in the SC class}.\textbf{--}
For a single electron in $SC(n,m,g^*)$ lattices, the energy spectra for Eq.~\ref{eq2} with zero on-site potentials are symmetric about 0 as a result of the bipartite lattice. We take the hopping strength $t$ as the energy unit and have $t=1$. In the local sites, the coordination number is 2, 3, or 4, so the bounded energy for all $SC(n,m,g^*)$ lattices is fixed from $-4t$ to 4t. In Fig.~\ref{Fig2:DireEne}(a1), \ref{Fig2:DireEne}(b1), \ref{Fig2:DireEne}(c1), and \ref{Fig2:DireEne}(d1), one can see the two traits mentioned above. Both features are independent of the $generator(n,m)$ and iterated patterns ($M_{se}$ and $M_{ge}$). At such a constricted energy span, the pattern remodulates the energy level density by the connection between the subblock lattices, the $generator(n,m)$ affects the level density by spatially local configurations (see four panels of Fig.~\ref{Fig2:DireEne}), and the role of three patterns is illustrated by the level distribution and DoS in Fig.~\ref{FigS1:DoS}. Remarkably, the tendency toward level attracting behavior becomes distinct, which is seen by enormous degenerate or quasidegenerate sub-clusters~\cite{Degeneratenote,Yao2022Wave}, as shown in left panels of Fig.~\ref{Fig2:DireEne}. This ``level clustering" behavior is rooted in fractal lattice topology, and intensified by increasing hierarchical level $g$, which is notable in the trends of the scattered points.

Furthermore, we analyze the bandwidth behavior of these planar $SC(n,m,g^*)$ lattice-based TB models, using four $generator(n,m)$ of $(5,3)$, $(4,2)$, $(3,1)$, and $(5,1)$ in Table~\ref{table:DH}. We plot the forbidden and allowed energies as a function of the geometrical hierarchical level $g^*$ and two patterns ($M_{se}$ and $M_{ge}$), see Fig.~\ref{Fig2:DireEne}(a1), \ref{Fig2:DireEne}(b1), \ref{Fig2:DireEne}(c1), and \ref{Fig2:DireEne}(d1). As in Fig.~\ref{FigS1:DoS}, the level legs are transformed to scatter points for better visibility. The allowed energy regions broaden for more significant $g^*$ in any patterns until they approach the upper limit of $4t$.

When we fix the pattern and adjust the graphical hierarchy level $g^*$ for each $generator(n,m)$, it is possible to identify the pattern effect. The bandwidth $2E_m$ or the upper bound energy $E_m$ is derived from Fig.~\ref{Fig2:DireEne}, and listed in Table~\ref{table:DH}. The results show that the $self$ pattern has the advantage of lengthening the bandwidth, unless the two patterns intersect coincidentally, such as in the $generator(3,1)$ case (see Fig.~\ref{Fig2:DireEne}(c1)). Given that the four $generator$s in Table~\ref{table:DH} have $n^2-m^2 \geq \mathcal{N}$ (here $\mathcal{N} = 8$), when  $g^*$ increases by a unit, the vertex network under the $gene$ pattern must proliferate more quickly than the self pattern. We conclude that the pattern option affects the bandwidth, with the $self$ pattern being more prominent.

Intriguingly, in Fig.~\ref{Fig2:DireEne}(d1) of the $generator(5,1)$ example, the bandwidth approaches the upper energy of $4t$ after $g^*$ reaches 2 or 3. In addition, far from the band center, it changes scarcely other than perhaps intensifying the degenerate behavior when $g^* \geq 3$. This claim applies to the infinite $g^*$ instance.

To exploit these bandwidths and the local band traits, another work is mentioned about Anderson transition in a Fibonacci chain~\cite{Machida1986Quantum}. Machida and Fujita rescaled the raw bands into the absolute unit interval, and noticed a dynamic band-branching tendency with increasing disorder strength, which is strongly related to the level correlation between some neighbor levels and the overlap between their wavefunctions. This prompts us to ask the following question: Can one control an analogous phase transition by adjusting some lattice topological parameters~\cite{Mandelbrot1982fractal,Gefen1980Critical,Yu1988Numerical,Mandelbrot1982fractal,Suzuki1983Phase,Gefen1983Geometric}?

Although the possible $g^*$ and the $(n,m)$ scenarios are limited by the lattice size $Ar$ we can simulate, the trend of the bandwidth changes and the local band trait in our results give some ample guidance. At initial $g*=2$, the upper bandwidth $E_m$ is near or beyond $3t$, see Table~\ref{table:DH}. It is expanded to the upper energy of $4t$ by varying the $generator(n,m)$ and/or increasing $g*$; as a result, the relative ratio of the rest bandwidth that we can extend is approximately $25\%$. Furthermore, some numerical local band clusters are shown in Fig.~\ref{Fig2:DireEne}(a2), \ref{Fig2:DireEne}(b2), \ref{Fig2:DireEne}(c2), and \ref{Fig2:DireEne}(d2), with no visible differences in the band tendency as $g^*$ increases.
We thus conclude that under the $M_{se}$ and $M_{ge}$ patterns, varying both the $generator(n,m)$ and $g^*$ cannot cause a phase transition, unless $SC(n,m,g^*)$ fractals in the $gene$ pattern approximate the square periodic-lattices, see the anomalous $SC(5,1,3^*)$ lattice. We further study the possible phase that all states are in when a single electron roams upon these $SC(n,m,g^*)$ lattices.

As the local band-branching predicts in Refs~\cite{Bloch1929,Anderson1958Absence,HOWLAND1987Perturbation,Hiramoto1989New,Fujiwara1989Multifractal,Hisashi1992},
a localized phase indicates the dense-point spectra, and an extended phase has the absolutely continuous spectra. A phase is neither of these but resides in between, its spectra are singularly continuous, and its wavefunctions are critical. Note that, to avoid losing some details when the point set is so dense, we also plot the integrated DoS $IDoS(E)$$=\int_{-\infty}^{E} \rho(E) d E$ and consider the $E\geq 0$ branch with a tiny energy window of 0.001, whose vertical staircase-like behavior signifies the location of the main gaps. Combining the above criteria and $IDoS(E)$, see Fig.~\ref{Fig2:DireEne}. The energy consecutiveness property is singular, and schemes such as stair are obvious in $IDoS(E)$ for all $SC(n,m,g^*)$ lattices in Table~\ref{table:DH}, except the $SC(5,1,3)$ lattice under the $gene$ pattern.

We are now striving to understand why the electron stays in the CS upon these fractal lattices. Two known examples have the CS. One is at the disorder-induced mobility edge, where the correlation length is typically comparable to the lattice size; the other is in pure lattice frustration with disrupted structural symmetry. Although there is the long-range order of scaling symmetry in these $SC(n,m,g^*)$ lattices, Bloch (extended) waves cannot be scattered over long distances due to the absence of translational symmetry. Either by Anderson localization~\cite{Anderson1958Absence}, or by local geometric design of the vertex surrounding~\cite{Vidal1998AB,Vidal2001Disorder,Mukherjee2018Experimental}, or both, advising all or some states into a localized phase. The above assumption fails in $SC(n,m,g^*)$ lattices because of the lack of disorder and no unique site network used for a single electron cage (here, both do not exist in our pure hopping model). Therefore, it makes sense for a mass state cluster to be in a critical phase, yet at the same time, these states are prone to some of the features of extended states~\cite{Lesser2022Emergence,Yao2022Wave}.

In Fig.~\ref{Fig2:DireEne}(d2), the $SC(5,1,3)$ lattice comes as a surprise, where the vertical stair-like traits disappear obviously when $g=3^*$ (see five insets $R1$-$R5$). One might assume that their wavefunctions are extended. In fractal reality, for a $generator(n,m)$ with large $n$ and small $m$, the proliferating lattices under the $gene$ pattern can be described as a translation-symmetry lattice with some point-like or cluster-like defects. Hence it is reasonable that there are possible densely continuous energy spectra when $g^*$ is significant. However, the above situation can be avoided in the $self$ pattern, the pattern $M_{se}$ always remains independent of the arbitrary $generator(n,m)$, and the scaling symmetry even becomes apparent by raising the $g^*$ value.
We next study the spectral correlation over short distances by level statistics $P(s)$ and $P(\tilde{r})$ in the RMT framework~\cite{dyson1963random,Guhr1998Random,Brody1981Random}.

\begin{figure*}[!htbp]
    \centering
    %  \subfloat[][]
    {\includegraphics[width=0.67\columnwidth]{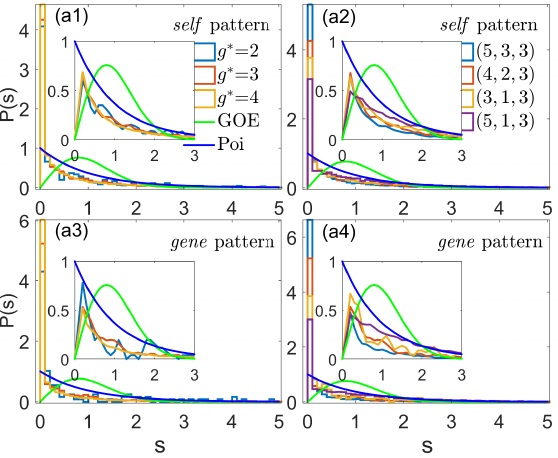}}
    {\includegraphics[width=0.67\columnwidth]{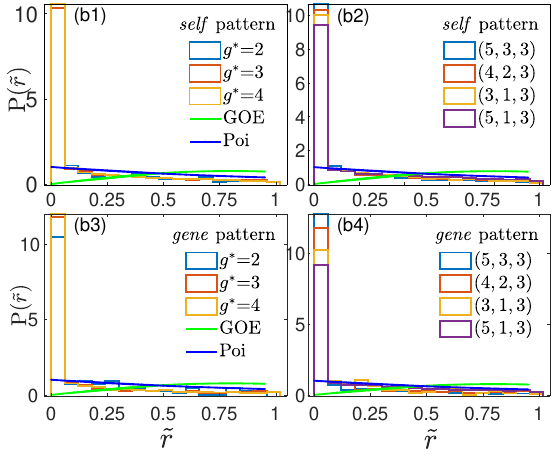}}
    {\includegraphics[width=0.67\columnwidth]{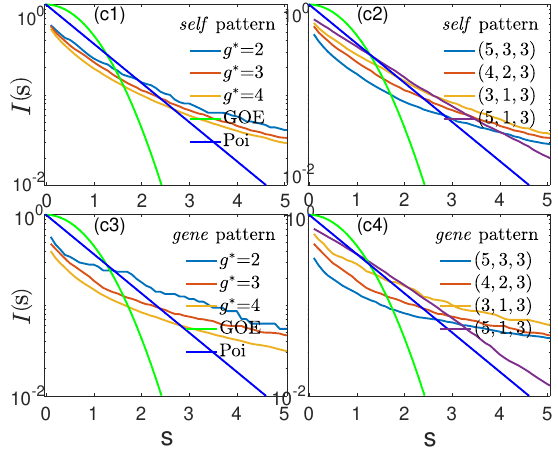}}
%      {\includegraphics[width=5.6cm,height=7.25cm]{Fig3Psnew.pdf}}
%     {\includegraphics[width=5.6cm,height=7.25cm]{Fig3Pr.pdf}}
%     {\includegraphics[width=5.6cm,height=7.25cm]{Fig3Is.pdf}}
    \caption{(Color online) Left panel: Level-spacing probability distribution histogram $P(s)$ after `unfolding' raw level sequences as a function of geometrical hierarchy level $g^*$ or a $generator(n,m)$. In (a1) and (a3), we consider three distinct $g^*=2, 3, 4$ cases with $(n,m)=(4,2)$ under the \textit{self} and \textit{gene} pattern, respectively; in (a2) and (a4), four $generator$s of $(5,3)$, $(3,1)$, $(4,2)$, $(5,1)$ in Table~\ref{table:DH} with fixed $g^*=3$ under the \textit{self} and the \textit{gene} pattern, respectively. The binwidth is 0.1. Middle panel: Gap-ratio probability distribution histogram $P(\tilde{r})$ benchmarks the raw levels, which corresponds to left panel. The binwidth is 0.06. Right panel: Integrated level-spacing distribution $I(s)$ has numerical stability in comparison with $P(s)$ and $P(\tilde{r})$. The linear-log scale shows $I(s)$ diversity near $s = 0$ between these above cases. The bindwidth is 0.001. The Poisson (solid blue line) and Wigner-surmise (solid green line) curves of Eqs.~\ref{eq3} and Eqs.~\ref{eq4} are included for reference. Numerical derivation $P(s)=-dI(s)/ds$ is added as four insets.}\label{Fig3:LevelSta}
\end{figure*}

\emph{Level correlation statistics in the SC class}.\textbf{--}
%\textit{Level correlation statistics in SC class.}--
For an obtained energy-level set \{$E_i, 1\leq i\leq Ar$ and $i$ is an integer\}, where the area value  $Ar = Dim(H)$. The DoS $\rho(E)$ divides into the glossy part $\rho_{gl}(E)$ and the local fluctuation $\rho_{fl}(E)$. $\rho_{gl}(E)$ needs to be extracted by analytical and numerical techniques. Unfortunately, this is not a simple task, especially given that the asymptotic one~\cite{Guhr1998Random,Torres2019Level} is difficult to obtain or causes some misleading results~\cite{Gomez2002Misleading}. We define a map $\varepsilon_i =\eta(E_i)$ at the original energy position $E_i$, which is obtained by interpolating the histogram $I(E)$ with a cubic spline~\cite{Zharekeshev1995Scaling} (further details are provided in App.~\ref{Sup:Unfolding}).

In contrast to $P(s)$ with $s_i=\varepsilon_{i+1}-\varepsilon_i$, Oganesyan and Huse~\cite{Ogan2007Local} define the gap-ratio $\tilde{r}$=min$(\delta, 1/ \delta)$ with $\delta=s_{i+1}/s_{i}$ and $s_i=E_{i+1}-E_i$, and the distribution $P(\tilde{r})$ has two benefits~\cite{Olivier2022Probing,Lucas2020Complex}: i) estimating $\rho_{gl}(E)$ is unnecessary, and we can bypass the numerical unfolding procedures; ii) its average value $\langle \tilde{r} \rangle$ characterizes the phase where it resides. Below, we exploit both $P(s)$ and $P(\tilde{r})$ statistics to diagnose the lattice topology of these $SC(n,m,g^*)$ fractals, as well as the computed $\langle \tilde{r} \rangle$ for quantitatively detecting their phase.

In left and middle panels of Fig.~\ref{Fig3:LevelSta}, the degenerate levels are not eliminated in these histograms, unlike in Ref~\cite{Zhong1998Level}. It is an evident justification for how a particular $generator(n,m)$ or the rise of $g^*$ impacts the local band gatherings. First, we discuss the overall characteristics of $P(s)$, which demonstrates whether all states live in a critical phase. Here, $s$ is normalized by the mean level spacing $\langle s \rangle$, and $s$ is set in a sufficient range from 0 to 5. Regardless of the dilation pattern or the configuration of the $generator(n,m)$ and $g^*$, the histogram of $P(s)$ has a remarkable peak near $s=0$ in our numeric, which has been signed by the singular continuity of the energy spectra. Additionally, $P(s)$ is neither a Poisson function for the localized phase nor a simple Wigner distribution for the extended phase~\cite{Guhr1998Random} in Eqs.~\ref{eq3},
\begin{eqnarray}\label{eq3}
P_{Poi}(s) &=& \exp (-s),  \notag\\
P_{GOE}(s) &=& (\pi/2)s \exp (-\pi s^2/4).
\end{eqnarray}
Conversely, it has $c_1 s^ \beta$ at small $s$ and decays as the stretched exponential form $\exp (-c_2 s^\alpha$) in the large $s$ range. The constants $c_1$ and $c_2$ are known by normalizing $P(s)$ and unitizing the mean value of $s$. The index $(\beta,\alpha)$ is fitted numerically (see the work~\cite{Brody1973A,Casati1991Scaling,Aronov1994pis,Varga1995Shape} of Brody, Izrailev, and Aronov \textit{et al.}).

Similarly, taking the $\tilde{r}$-value as a metric, both standard distributions have equivalent expressions~\cite{Atas2013Distribution,De2022Level},
\begin{eqnarray}\label{eq4}
P_{Poi}(\tilde{r}) &=& \frac{2}{(1+\tilde{r})^2},  \notag\\
P_{GOE}(\tilde{r}) &=& \frac{27}{4}\frac{\tilde{r}+\tilde{r} ^2}{(1+ \tilde{r}+\tilde{r} ^2)^{5/2}}.
\end{eqnarray}
The mean value is $\langle \tilde{r}\rangle_{Poi}=0.386$ for the localized phase and $\langle \tilde{r}\rangle_{GoE}=0.536$ for the extended phase. As we know, these levels become severely (quasi-)degenerate when taking fairly large $g^*$, which causes $\langle \tilde{r} \rangle$ in our results to fall below 0.386, as demonstrated in Table~\ref{table:DH}.
To clarify how the local band gathering relates to the pattern and the $generator(n,m)$. We employ the identical tactic by fixing the $g^*$ value and comparing four $generator(n,m)$ in $D_H$ ascending order (see Table~\ref{table:DH}). The same $generator(n,m)$ sequences are also applied to the $self$ pattern.

First, focusing on the \textit{self} dilation pattern, we select the $generator(4,2)$ case as illustrated in Fig.~\ref{Fig3:LevelSta}(a1). At $s$ close to 0, the peak height of $P(s)$ grows with geometrical hierarchy level $g^*$, which is attributed to the enhanced level clustering effect. As $s$ moves away from 0, there are fewer differences between these LSDs $P(s)$, whereas they are generally below $P_{Poi}(s)$ (blue curve in Fig.~\ref{Fig3:LevelSta}) until $s$ approaches the crossover point of 2.0019 where Eqs.~\ref{eq3} meet. Note that, we numerically interpolate Eqs.~\ref{eq3} to obtain two crossover points of 1.2732 and 2.0019. As $s$ grows persistently beyond it, these LSDs would surpass the Wigner distribution of $P_{GOE}(s)$ (green curve in Fig.~\ref{Fig3:LevelSta}). These critical distributions have somewhat Wigner-like features; however, they differ from the critical behavior when Anderson localization occurs. The latter has two traits: one is the critical distribution retained between the extended and localized phases; the other is that it intersects obviously with $P_{Poi}(s)$ (see Fig. 1 in Ref~\cite{Zharekeshev1995Scaling} of Zharekeshev \textit{et al.}). We also know the impact of the $generator(n,m)$ on $P(s)$, whose values are taken from Table~\ref{table:DH} with fixing $g^*=3$. When $s$ is near 0, the peak height of $P(s)$ is lessened for the sequences of $(5,3)$, $(3,1)$, $(4,2)$, and $(5,1)$. As we know, $\mathcal{D}_{se}$ is always the same when taking an enormous $g^*$ value, regardless of the $generator(n,m)$. We suspect that pre-fractal $SC(n,m,g^*\ll\infty)$ accounts for it. These above arguments can be verified again by alternative spectra $P(\tilde{r})$ (also see in Fig.~\ref{Fig3:LevelSta}(b1) and \ref{Fig3:LevelSta}(b2), whose trends of the histograms are consistent with the curves of Fig.~\ref{Fig3:LevelSta}(a1) and \ref{Fig3:LevelSta}(a2), respectively).

Now we turn to the \textit{gene} pattern. Intuitively, for the $P(s)$ histograms in Fig.~\ref{Fig3:LevelSta}(a3) and \ref{Fig3:LevelSta}(a4) and the $P(\tilde{r})$ histograms in Fig.~\ref{Fig3:LevelSta}(b3) and \ref{Fig3:LevelSta}(b4), the critical phase is still distinct from those of the known Anderson transition. In addition, two comparable hallmarks about the peak height of $P(s)$ in the small-$s$ region can be pictured: i) if $g^*$ is increased, the level clustering behavior obtains a boost for the two patterns we considered, depicted in Fig.~\ref{Fig3:LevelSta}(a1) and \ref{Fig3:LevelSta}(a3). In the $gene$ pattern, the peak height is higher than the appearance in the $self$ pattern. Consequently, the dilation pattern plays a role in making level attraction ii) And conversely, when $g^*$ remains unchanged, the $generator(n,m)$ is arranged in ascending $\mathcal{D}_H$ order, and by increasing the value $n^2-m^2$, the peak value of $P(s)$ gradually drops, see Fig.~\ref{Fig3:LevelSta}(a4).

\begin{figure}[!htbp]
%\centering
  {\includegraphics[width=8.0cm,height=8.0cm]{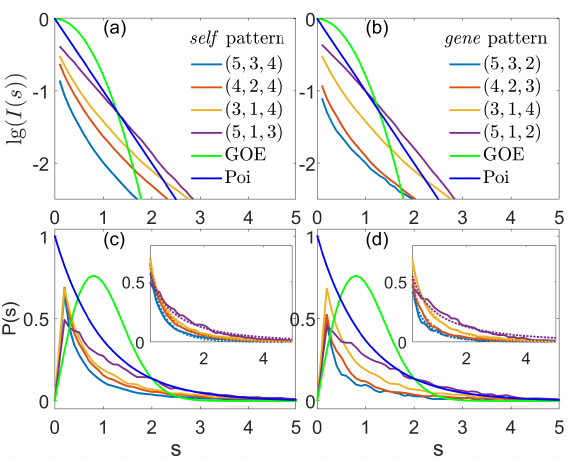}}
 % {\includegraphics[width=3.4in]{Fig4IsPsScaling.pdf}}
  \caption{(Color online) Finite-size scaling of $I(s)$ and $P(s)$ for two \textit{Sierpi\'{n}ski carpet} subclasses, which are constructed by Eq.~\ref{eq1} with four $generator$s in Table~\ref{table:DH}. Under the $self$ ($gene$) pattern, the lattice size $Ar$ is 65536 (4096), 49152 (20736), 32768 (32768), and 12288 (13824) from top to bottom, here we use the hierarchy level $g$. The linear-log scale $I(s)$ is in (a) and (b). Numerical derivation $P(s)=-dI(s)/ds$ are shown in (c) and (d). The two insets illustrate the fitting performance (dashed line) for the asymptotic tail.}\label{Fig4:logIs}
\end{figure}

Thus far, we can infer that when a single electron hops purely in these $SC(n,m,g^*)$ lattices, their LSDs tend to have a larger fraction in the small-$s$ domain with increasing iteration depth $g^*$ or Hausdorff dimension $\mathcal{D}_{ge}$, which suggests designing many fractal-shape artifacts~\cite{Kopelman1997Spectroscopic,Hernando2015Spectral} with some functions of absorbing continuous optical spectra.
Since the histograms of $P(s)$ and $P(\tilde{r})$ in our results scale dependently on the different binwidths, to quantitatively distinguish these $P(s)$, we adjust with a stabilized integrated LSD $I(s)=\int_s^\infty P(t) dt$, which counts the fraction of the level spacing that is not less than $s$, and hence $I(0)=1$.
It is simpler to grasp the level correlation behavior in a small or large $s$ range, also integrating Eqs.~\ref{eq3} as two references, both yield the gestures $I_{Poi} =\exp (-s)$ and $I_{GOE} =\exp (-\pi s^2/4)$, respectively.

Considering that $P(s)>P_{GOE}(s)$ in the large $s$ region, this diversity is highlighted by simply mapping it to $I(s)$, see right panel of Fig.~\ref{Fig3:LevelSta}. We notice that the huge peak of $P(s)$ causes $I(s)$ to climb quickly at small $s$ positions,
which is attributed to many degenerated and quasidegenerated levels in $SC(n,m,g^*)$ lattices. We strip these degenerated behaviors by $P(s)=-dI(s)/ds$; hence, the dramatic peak is below that of the histograms (the four insets in left panel of Fig.~\ref{Fig3:LevelSta}); notably, this peak is distinct from that in Anderson transition. In the latter case, the corresponding states are critical near the mobility edge drawing from the localized phase to the extended phase or vice versa. Still using $P(s)$ terminology, there is a cross point of $s=1.2732$ between $P_{Poi}(s)$ and $P_{GOE}(s)$; the anticipated critical distribution~\cite{Zharekeshev1995Scaling} also bisects with $P_{Poi}(s)$, in which the crossover point is less than 1.2732 and decreases gradually when the disorder strength increases. However, the above characteristics differ from $P(s)$ associated with $SC(n,m,g^*)$ lattices. $P(s)$ in our results indeed belongs to a new critical phase, and it is typically caused by fractal lattice topology.
To strengthen the above arguments, we perform a finite-size scaling analysis. In Fig.~\ref{Fig4:logIs}, for each $generator(n,m)$ with the maximally computed $g^*$ under two patterns (lattice size $Ar\sim 10^4$ in most $SC(n,m,g)$ lattices), the LSDs still exhibit the same traits.

\begin{table}%[!htbp]%[H]
\begin{ruledtabular}
\caption{\label{table:II} Fig.~\ref{Fig4:logIs}(c) and (d) provide the fitting parameter cell $(c_1, c_2, \alpha)^T$, and $g$ takes the same value in (a) and (b). The transcendental function nature of potential $P_{fit}(s)=c_1 s \exp(-c_2 s^\alpha)$ causes the inaccessibility of $(c_1, c_2$) by normalizing $P(s)$ and unitizing $\langle s\rangle$ that we mentioned previously. In general, $\alpha$ cannot be known analytically. As an  alternative, we select the tiny $s$-region $[0,5]$ and piecewise fit $P(s)=-dI(s)/ds$: linear formula $c_1 s$ for increasing curve and asymptotic tail $\exp(-c_2 s^\alpha)$ for decay curve (see two insets in Fig.~\ref{Fig4:logIs}(c) and (d)) with a $95\%$ confidence level.}
\begin{tabular}{p{0.2\linewidth}p{0.17\linewidth}p{0.16\linewidth}p{0.16\linewidth}p{0.16\linewidth}}
        $SC(n,m,g)$ & (5, 3) & (4, 2) & (3, 1) & (5, 1) \\ \hline
        \multirow{3}{*}{$self$} & (3.261, & (3.462 & (3.414  & (2.458, \\
                              & \ 2.103, & \ 1.751 & \ 1.600  & \ 1.434,  \\
                              & \ 0.818) & \ 0.797) & \ 0.768) &\ 0.584) \\ \hline
        \multirow{3}{*}{$gene$} & (2.378, &  (2.634, & (3.414, & (2.248, \\
                              & \ 2.471, & \ 2.146, & \ 1.600, & \ 1.426, \\
                              & \ 0.641) & \ 0.668) & \ 0.768) & \ 0.539) \\
\end{tabular}
\end{ruledtabular}
\end{table}

One may notice that $P(s)$ and $I(s)$ act on a scale (four insets in Fig.~\ref{Fig3:LevelSta}, Fig.~\ref{Fig3:LevelSta}(c), and Fig.~\ref{Fig4:logIs}). $P(s)$ cannot be fitted effectively due to the irregular fluctuation at its tail. In addition, when unfolding the raw levels, the peak height of $P(s)$ and its location depend on the binwidth of histograms. We further discuss the Winger-like behaviors of $P(s)$ using eight $SC(n,m,g^*)$ lattices in Fig.~\ref{Fig4:logIs}, where $Ar$ is the upper size we can simulate. First, the linear trait is apparent as $s$ approaches 0, such as $P_{GOE}(s\rightarrow 0)\propto s$, whose models are real and symmetric matrices, so both yield identical orthogonal ensembles ($\beta=1$). There is a more notable slope as $Ar$ grows due to the augmented level clustering property. Second, the tail asymptotic $P(s) \propto \exp(-c_2 s^\alpha)$ at $s$ exceeds the value pinned by the peak of $P(s)$ but is less than 1, as shown in Fig.~\ref{Fig4:logIs}(c) and \ref{Fig4:logIs}(d), has an intermediate power law decay, as proven analytically in Ref~\cite{Askar2019Power}. If we exploit a shape-preserving formula $P_{fit}(s)=c_1 s \exp(-c_2 s^\alpha$) to fit these $P(s)$, that is a challenge due to the shaking tails when the data $P(s\geq 5)$ are considered. We instead use two piecewise functions and obtain three parameters $(\alpha,c_1,c_2)$, attached in Table~\ref{table:II}. The fitting performance is shown in Fig.~\ref{Fig4:logIs}(c) and \ref{Fig4:logIs}(d) insets.
%%%%%%%%%%%%%%%%%%%%%%%%%%%%%%%%%%%%%%%%%%%%%%%%%%%%%%%%%%%%%%%%%%%%%%%%%%%%%%%%%%%%%%%%%%%%%%%%%%

\section{Conclusion}\label{Conclu}
In conclusion, for a noninteracting electron gas confined in two $SC(n,m,g^*)$ lattice classes, percolated separately by the $self$ and $gene$ patterns, we have analyzed their electronic energy spectra, and have confirmed that electronic states are embedded in a critical phase, by singularly continuous spectra and critical level-spacing distribution $P(s)$ and level-spacing ratio distribution $P(\tilde{r})$. Although we enumerate several $SC(n,m,g^*)$ lattices in Table~\ref{table:DH}, for more cases with larger $g^*$, we can still exploit the tendency of the bandwidth change and local band trait to forecast that they belong to the same critical phase. Even by tuning the $generator(n,m)$ in the $SC(n,m,g^*)$ lattices, such a critical phase cannot be transited in other phases of the extended or localized phase. The critical phase for these fractals results from broken translation symmetry and the long-range order of scaling symmetry. We note that $M_{ge}$ is always determined by their $generator(n,m)$; in some extreme cases where $n$ is relatively large compared to $m$, the dilated lattice in the $gene$ pattern approximates roughly as a 2D periodic lattice with some point or small-cluster defects, which causes somewhat extended traits in its energy spectra. The critical phase exists in these $SC(n,m,g^*)$ lattices, unlike the well-known extended and localized phases and that near mobility edge when (Anderson) phase transition occurs. In addition, an increase in the $g^*$ value demonstrates that the level attraction or clustering exhibited in these lattices is due to scaling symmetry.

We also note that the renormalization group approach was used in 1D Fibonacci chains, 2D lattices such as Vicsek~\cite{You1993Exact} or other fractals~\cite{Domany1983Solutions,Iliasov2020Linearized}, and Penrose tiling~\cite{You1992Local}. However, due to the large lattice size, the decimating procedures become very tedious in the $SC(n,m,g^*)$ lattices. As an alternative, multifractal analysis~\cite{Yao2022Wave} is preferred for these critical states. A simple hopping model is exploited to explore the $SC(n,m,g^*)$ fractal topology. Still, it is generalized easily to the disordered Anderson model~\cite{Anderson1958Absence} and its variations~\cite{Harper1955Single,Aubry1980analyticity,Harper1955General} or includes the two-body interaction~\cite{De2022Level}. Alternatively, other particles such as phonon are in quasicrystals~\cite{Salazar2003Phonon} and fractals.

\par We appreciate some comments from Achille Mauri and Tjacco Koskamp on this project. This work was supported by the National Natural Science Foundation of China (Grant No. 12174291) and the National Key R\&D Program of China (Grant No. 2018YFA0305800) and computational resources from the Supercomputing Center of Wuhan University. Q.Y. acknowledges the grant support by the China Scholarship Council (CSC) through file No. 202006270212 when attending Radboud University in the Netherlands. X.Y. is supported by China Scholarship Council (CSC) under Grants No. 202106270050.
\begin{appendix}
\setcounter{figure}{0}
\renewcommand{\thefigure}{S\arabic{figure}}
\vspace{10 pt}

\begin{figure*}[!htbp]%[!htp]
    \centering
    %  \subfloat[][]
      {\includegraphics[width=0.67\columnwidth]{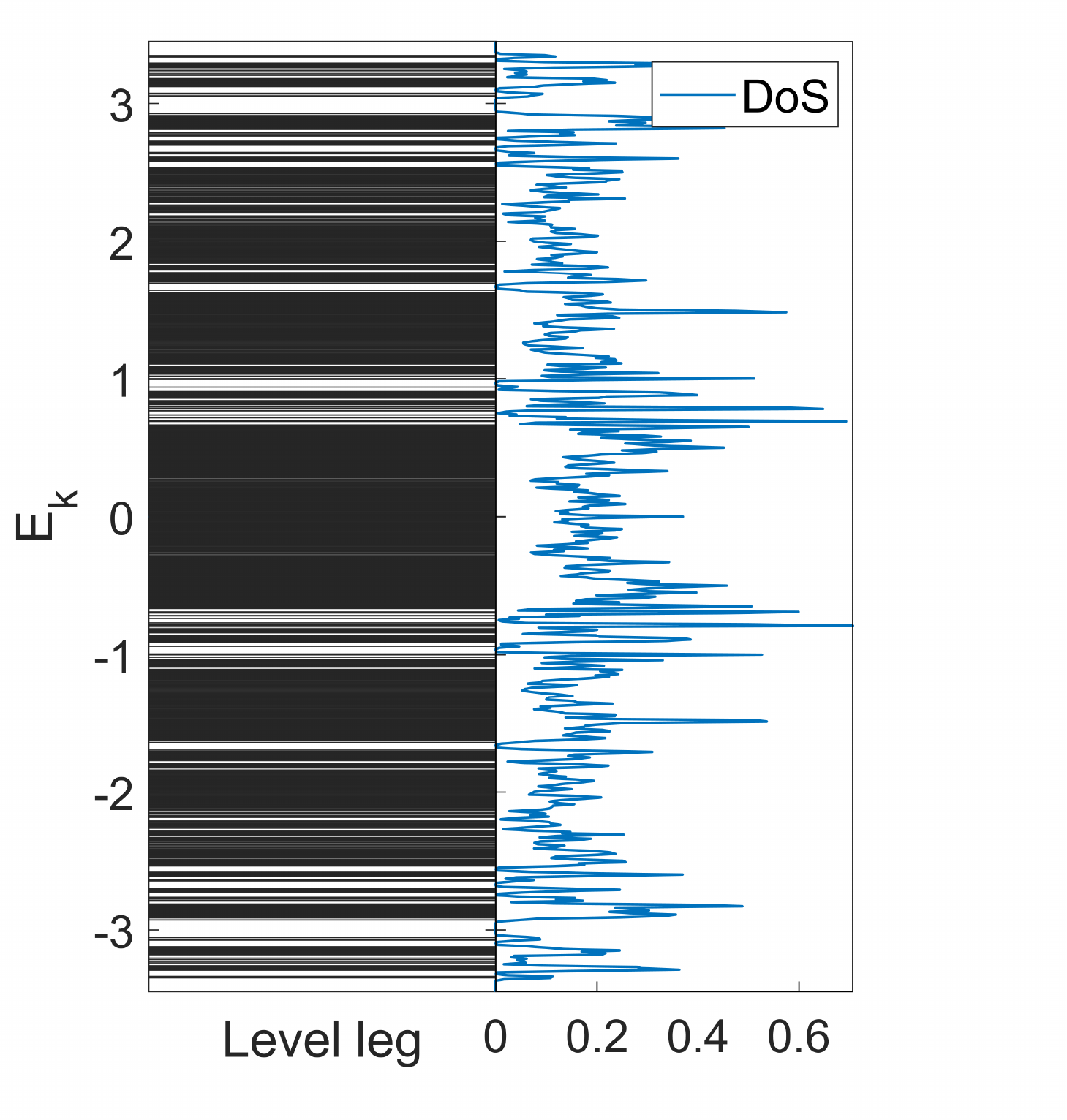}}
     %\subfloat[][]
     {\includegraphics[width=0.67\columnwidth]{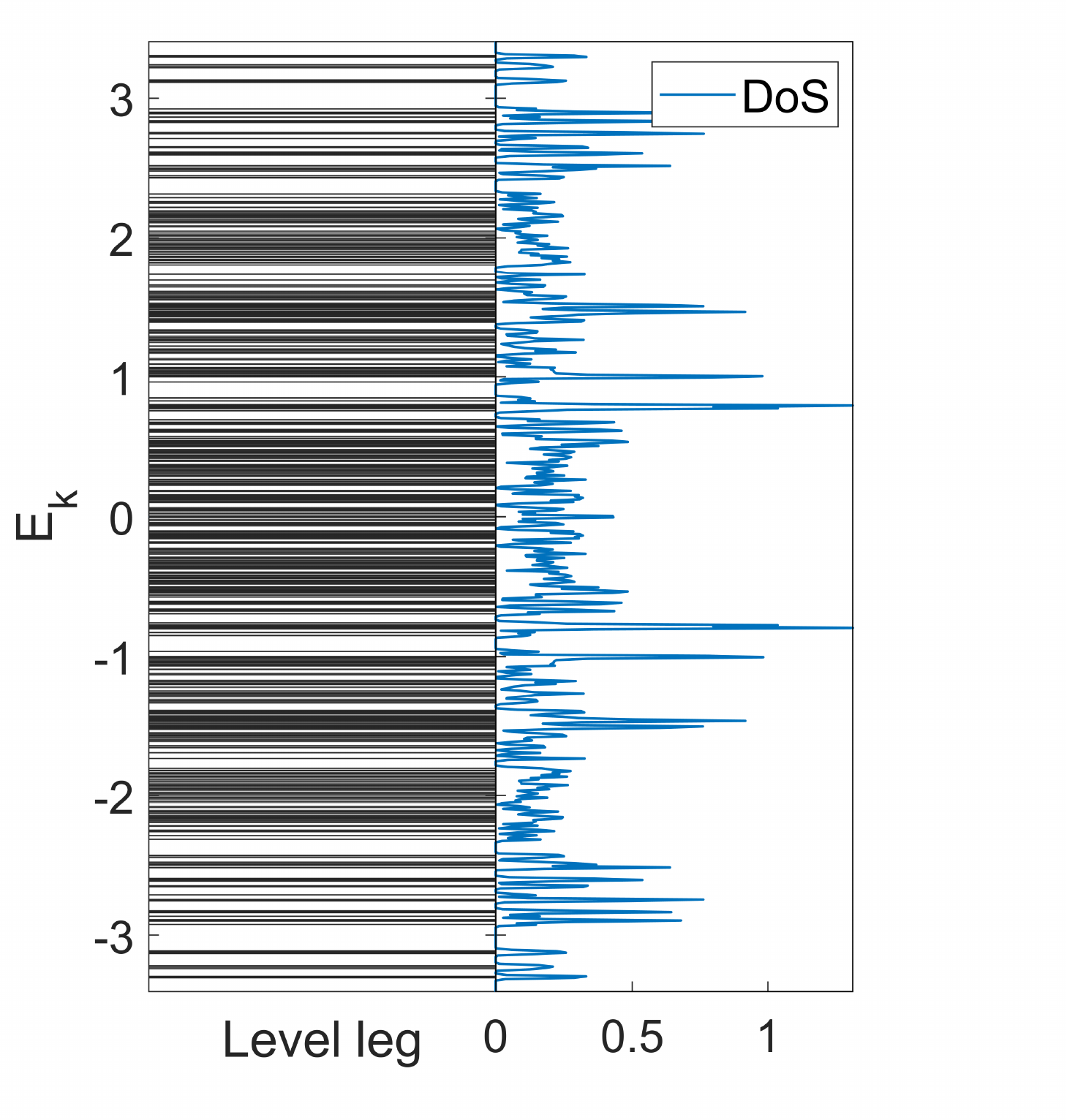}}
     {\includegraphics[width=0.67\columnwidth]{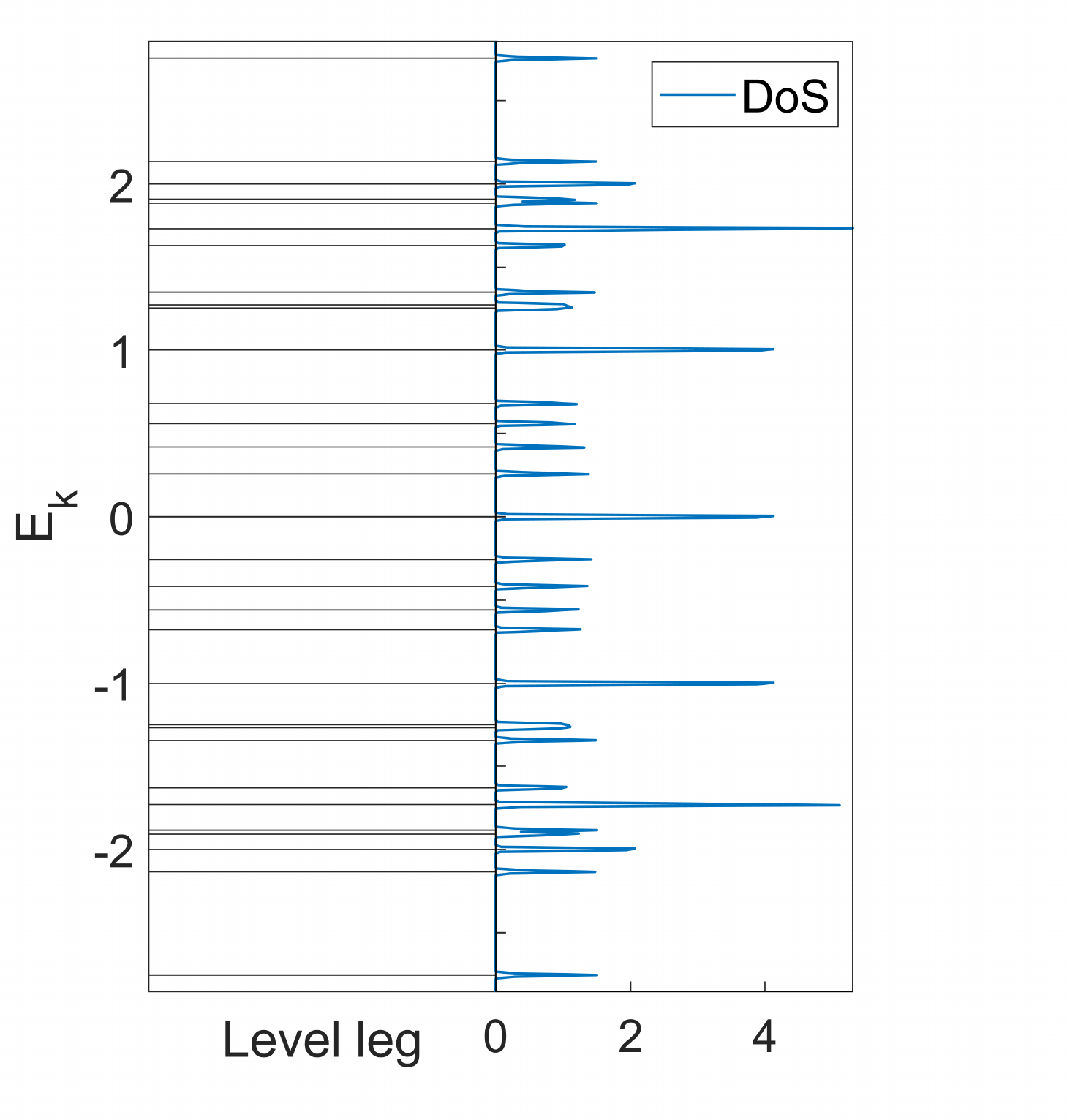}}
    \caption{(Color online) Level leg and density of state (DoS) for the tight-binding model upon the corresponding fractal $SC(4,2,2)$ lattices that are produced under three patterns. From left to right panels: the $self$, the $gene$, and the variant of the $self$ pattern (or called the $vari$ pattern).}\label{FigS1:DoS}
\end{figure*}

\begin{figure*}[!htbp]
    \centering
    %  \subfloat[][]
    {\includegraphics[width=0.67\columnwidth]{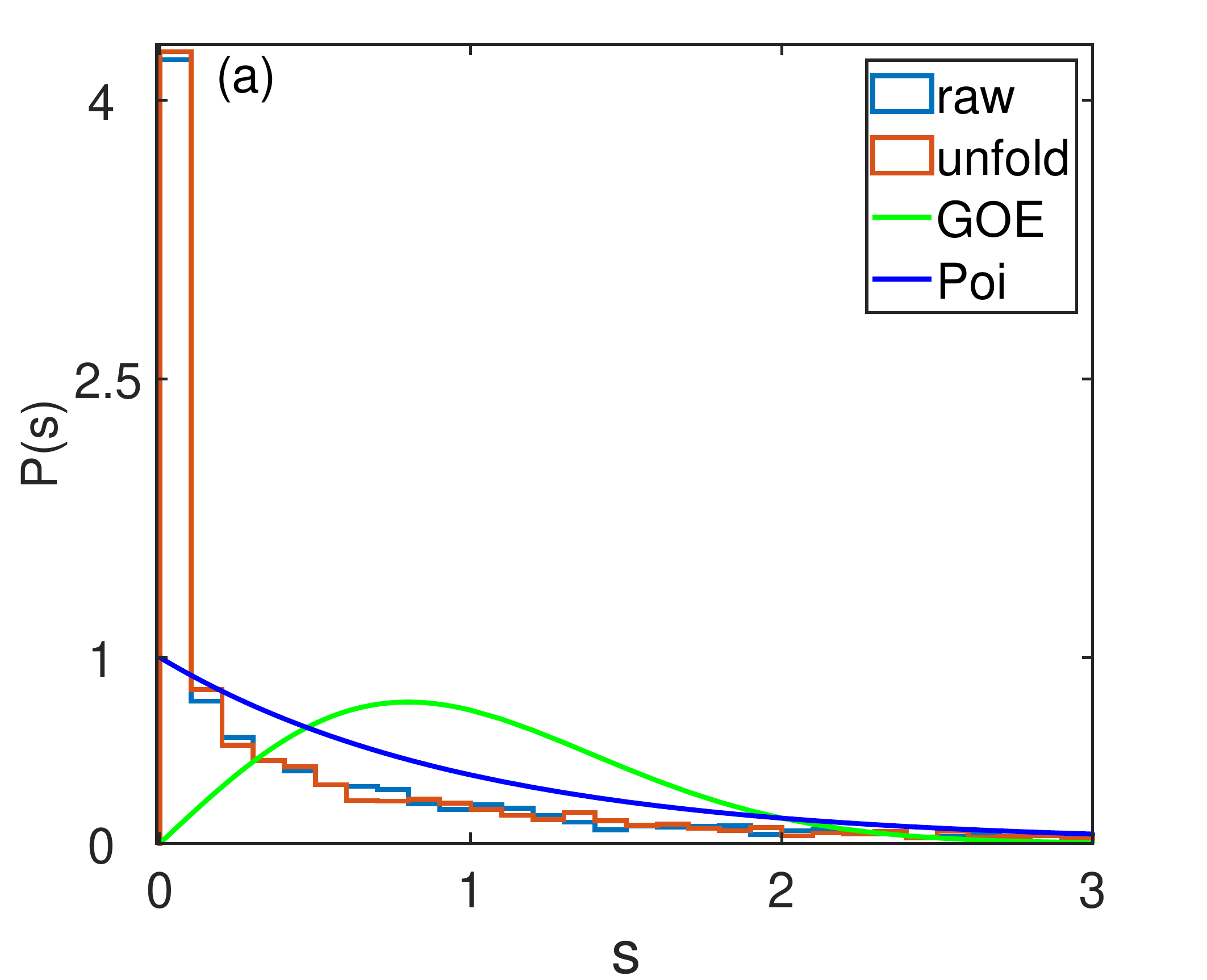}}
    {\includegraphics[width=0.67\columnwidth]{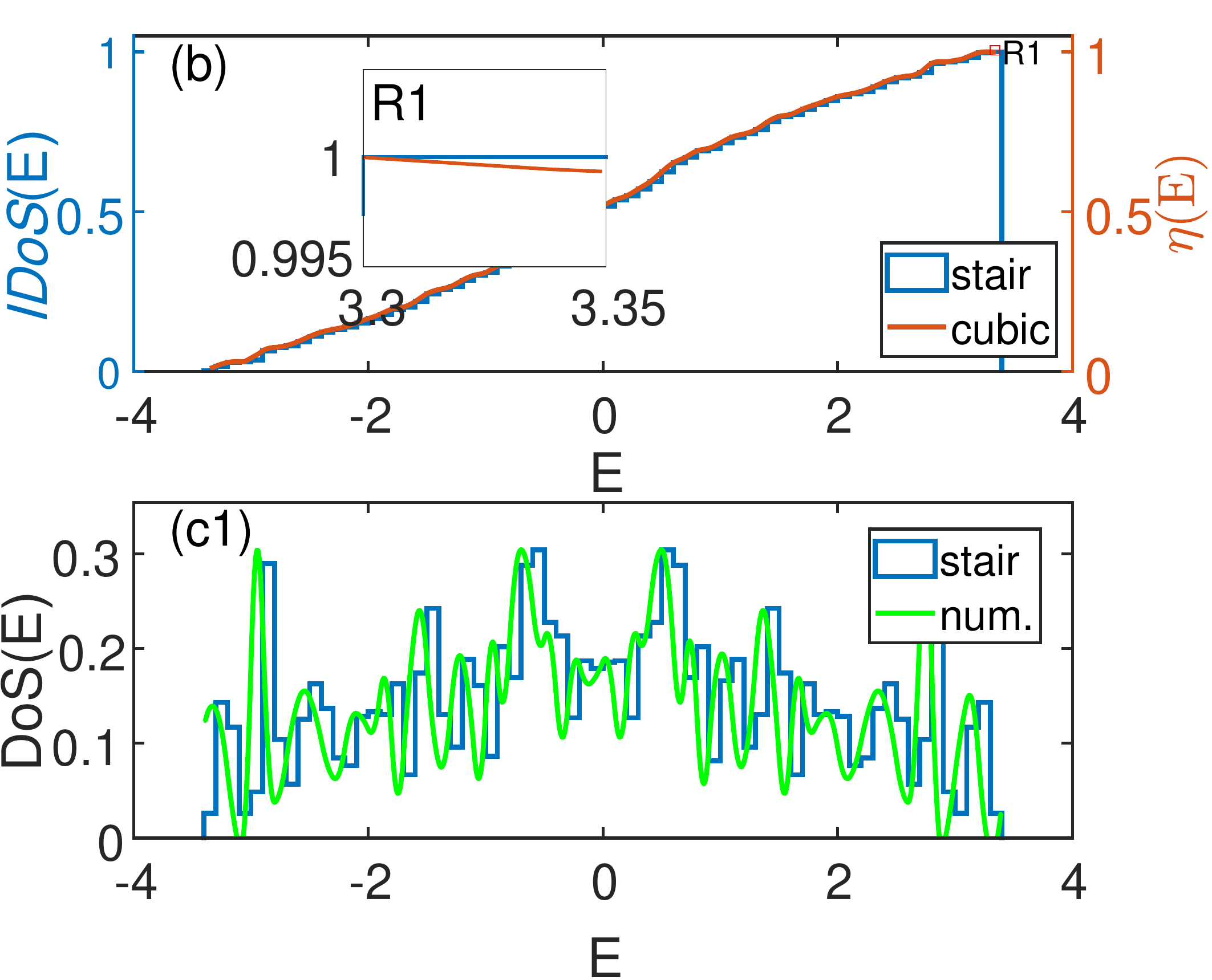}}
    {\includegraphics[width=0.67\columnwidth]{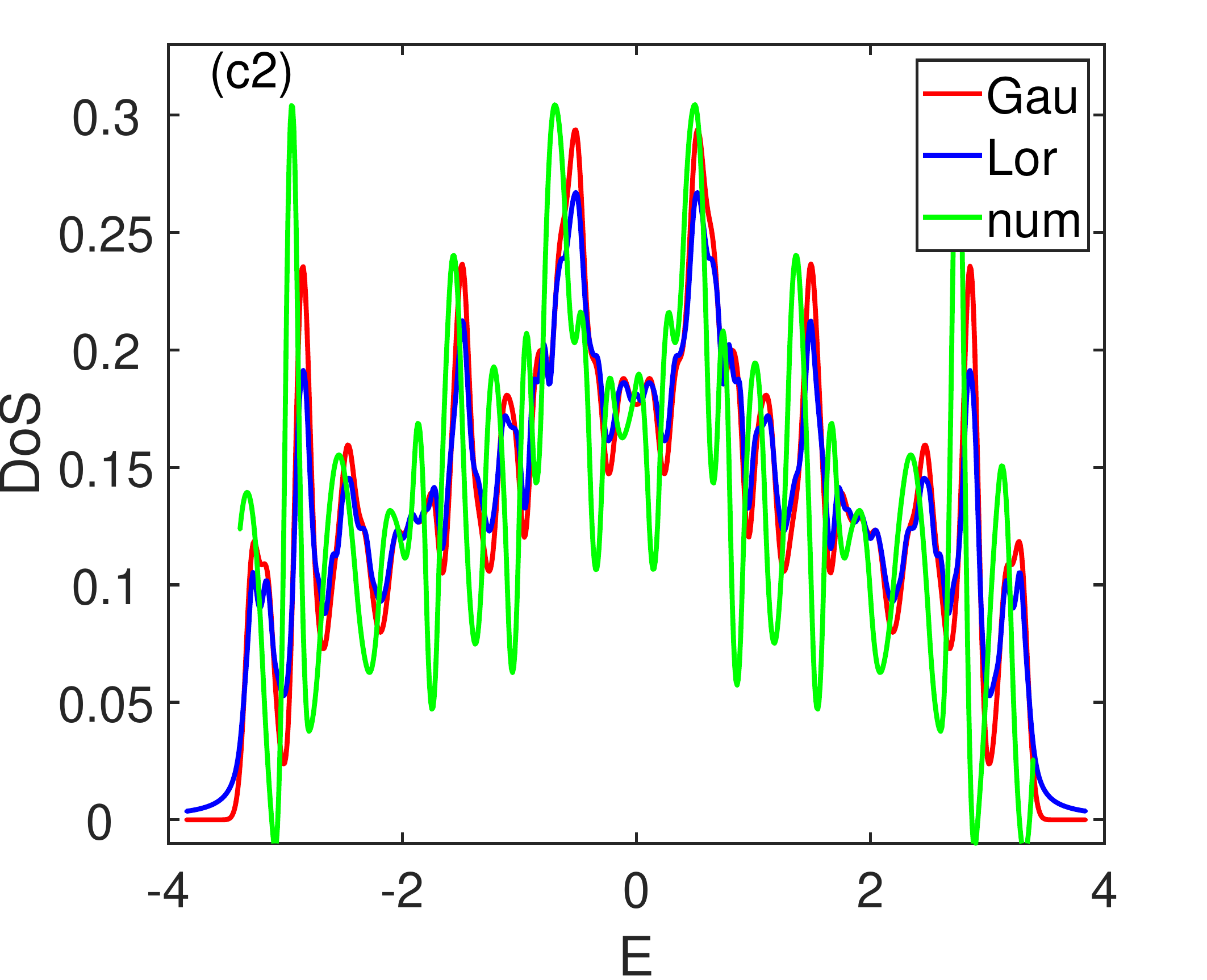}}
    \caption{(Color online) $P(s)$ after and before the unfolding procedure, are shown in (a), adding $P_{GOE}(s)$ and $P_{Poi}(s)$ as a comparison. $P(s)$ is derived indirectly from the cubic interpolation function $\eta(E)$, and $\eta(E)$ is obtained from integral DoS $IDoS(E)$ in (b). The monotonicity of $\eta(E)$ is violated in some region (red rectangle $R1$ is zoomed out as an inset). In (c1) DoS is the derivative of $\eta(E)$ (solid green), and we compare it with the original histogram (stair blue). In (c2), the Gaussian (solid red) or Lorentz (solid blue) function $\rho_{\sigma}(E)$ regularizes the needle-like energy spectra $\rho(E)=\sum_n\delta(E-E_n)$, and a tiny energy window is $\sigma=0.05$. Both are compared with $\rho(E)=d\eta(E)/dE$ (solid green).}\label{FigS2:Unfolding}
\end{figure*}

We supplement the analysis of the main text in two appendices: (i) the discrete level distributions are provided, together with their regularized density of state; (ii) the numerical unfolding technique for raw level sequences $\{E_1\leq E_2<...<E_k..., k=3, 4, ..., \textit{Dim}(H)\}$ is declared, these new levels are then employed for $P(s)$ statistics.
%In these appendices below, we add some detail to the discussion of the main text: (i) the discrete level distributions and their regularized density of state are presented; (ii) numerical unfolding procedures for raw level sequences $\{E_1\leq E_2<...<E_k..., k=3, 4, ..., \textit{Dim}(H)\}$ are demonstrated, and then these modified levels are used for $\mathbf{s}$-\emph{}value $P(s)$ statistics.%; (iii) some comments about long-range correlation between remote levels are discussed.

\section{level-distribution density}\label{Sup:Density}
We describe a single electron hops purely in three $SC(n,m,g^*)$ lattices by the tight-binding model. Three lattices are constructed under three patterns of the $self$, $gene$, and $vari$ patterns. The former two patterns are our focus, and note that the $vari$ pattern is a variation of the $self$ pattern, which can be handled from the possible $M_{se}$. After direct diagonalization, the raw levels are obtained and are shown in level leg and DoS. The Dirac function is blurred under some approximations to avoid the needlelike peaks in DoS. We can choose either an advanced Gaussian or Lorentz function~\cite{Lin2016approximating,Yao2022Wave} for this approximation. The tiny energy window is $0.005$.

As we compare three pattern-based lattices, we set the same geometrical hierarchy $g^*$. Moreover, if $g^*$ is small, the local nearby cluster gets obvious. Hence we have $g^*=2$. In Fig.~\ref{FigS1:DoS}, the prominent symmetry about $E=0$ is marked. In the $vari$ pattern of an example, $M_{va}=[1\;0\;0; 0\;0\;1; 1\;1\;0]$, some disconnected vertex-subclusters in $SC(n,m,g^*)$ lattices cause some subblocks in the \textit{Hilbert space}. Eventually, the lattice topology shows up by splitting its spectra into a few typical but small clusters; it is evident that increasing $g^*$ can boost its peak value but cannot alter its locations where the peaks are pinned (see right panel of Fig.~\ref{FigS1:DoS}).

In contrast, the DoS in the $self$ and $gene$ patterns is coherent and dense (the left and middle panels in Fig.~\ref{FigS1:DoS}), which is attributed to the absence of discrete vortex-subclusters in their lattices. Meanwhile, the uniform trend can be observed in two patterns. In the $gene$ pattern, the level attracting is more intense near the band center, and the $self$ pattern has a blocky spectrum. Their diversities are ascribed to the pattern dilation effect. Note that the initial $g^*=2$ case for the $self$ pattern has more vertexes, hence it appears to have more level legs. In the main text, we have elaborated on some discussions concerning the $generator(n,m)$ and $g^*$.

\section{Unfolding procedures for level-spacing statistic}\label{Sup:Unfolding}
The spirit of the unfolding procedure has been declared in the second subsection of Sec.~\ref{FraLS}, for the integrated DoS $IDoS(E)$$=1/Ar \sum_{m} \Theta\left(E-E_{m}\right)$, where $\Theta(t)$ is a cumulative count function for $t\leq 0$, we eliminate the discontinuity of $IDoS(E)$ by cubic spline and obtain $\eta(E)$, shown in Fig.~\ref{FigS2:Unfolding}(b). Note that, less than $5\%$ of numerical points violate the monotonicity of $IDoS(E)$ (see the inset $R1$). We skip them and link the following points to keep $\eta(E)$ monotonous.

Then, using two different methods, we compare the derivative $\rho(E)=d\eta(E)/dE$ to ensure the correctness of $\eta(E)$. One is by a DoS that resembles a histogram in Fig.~\ref{FigS2:Unfolding}(c1). Our results confirm that both are primarily the same, and the horizontal offset is caused by the forward Euler method. The other is by a smooth-like DoS in Fig.~\ref{FigS2:Unfolding}(c2), which exploits the same Gaussian or Lorentz blurring function. Similarly, the regularized DoS $\rho_{\sigma}(E)$ with these two functions caters to the envelope of numerical derivative $\rho(E)$ (note that the blurring window and the derivative interval are not the same, hence the requirement of completely coincident curves is not necessary.). To  obtain the new level ones $\{\varepsilon_i\}$ for $P(s)$, we interpolate the $\varepsilon_i=\eta(E_i)$ at raw level $E_i$. In Fig.~\ref{FigS2:Unfolding}(a), using an illustrated $SC(4,2,2)$ lattice, $P(s)$ is shown with the raw and new level sequences, respectively.
\end{appendix}

%\bibliography{ref}

\begin{thebibliography}{71}%
\makeatletter
\providecommand \@ifxundefined [1]{%
 \@ifx{#1\undefined}
}%
\providecommand \@ifnum [1]{%
 \ifnum #1\expandafter \@firstoftwo
 \else \expandafter \@secondoftwo
 \fi
}%
\providecommand \@ifx [1]{%
 \ifx #1\expandafter \@firstoftwo
 \else \expandafter \@secondoftwo
 \fi
}%
\providecommand \natexlab [1]{#1}%
\providecommand \enquote  [1]{``#1''}%
\providecommand \bibnamefont  [1]{#1}%
\providecommand \bibfnamefont [1]{#1}%
\providecommand \citenamefont [1]{#1}%
\providecommand \href@noop [0]{\@secondoftwo}%
\providecommand \href [0]{\begingroup \@sanitize@url \@href}%
\providecommand \@href[1]{\@@startlink{#1}\@@href}%
\providecommand \@@href[1]{\endgroup#1\@@endlink}%
\providecommand \@sanitize@url [0]{\catcode `\\12\catcode `\$12\catcode
  `\&12\catcode `\#12\catcode `\^12\catcode `\_12\catcode `\%12\relax}%
\providecommand \@@startlink[1]{}%
\providecommand \@@endlink[0]{}%
\providecommand \url  [0]{\begingroup\@sanitize@url \@url }%
\providecommand \@url [1]{\endgroup\@href {#1}{\urlprefix }}%
\providecommand \urlprefix  [0]{URL }%
\providecommand \Eprint [0]{\href }%
\providecommand \doibase [0]{https://doi.org/}%
\providecommand \selectlanguage [0]{\@gobble}%
\providecommand \bibinfo  [0]{\@secondoftwo}%
\providecommand \bibfield  [0]{\@secondoftwo}%
\providecommand \translation [1]{[#1]}%
\providecommand \BibitemOpen [0]{}%
\providecommand \bibitemStop [0]{}%
\providecommand \bibitemNoStop [0]{.\EOS\space}%
\providecommand \EOS [0]{\spacefactor3000\relax}%
\providecommand \BibitemShut  [1]{\csname bibitem#1\endcsname}%
\let\auto@bib@innerbib\@empty
%</preamble>
\bibitem [{\citenamefont {Zhong}\ \emph {et~al.}(1998)\citenamefont {Zhong},
  \citenamefont {Grimm}, \citenamefont {R\"omer},\ and\ \citenamefont
  {Schreiber}}]{Zhong1998Level}%
  \BibitemOpen
  \bibfield  {author} {\bibinfo {author} {\bibfnamefont {J.~X.}\ \bibnamefont
  {Zhong}}, \bibinfo {author} {\bibfnamefont {U.}~\bibnamefont {Grimm}},
  \bibinfo {author} {\bibfnamefont {R.~A.}\ \bibnamefont {R\"omer}},\ and\
  \bibinfo {author} {\bibfnamefont {M.}~\bibnamefont {Schreiber}},\ }\bibfield
  {title} {\bibinfo {title} {Level-spacing distributions of planar
  quasiperiodic tight-binding models},\ }\href
  {https://doi.org/10.1103/PhysRevLett.80.3996} {\bibfield  {journal} {\bibinfo
   {journal} {Phys. Rev. Lett.}\ }\textbf {\bibinfo {volume} {80}},\ \bibinfo
  {pages} {3996} (\bibinfo {year} {1998})}\BibitemShut {NoStop}%
\bibitem [{\citenamefont {Mandelbrot}(1982)}]{Mandelbrot1982fractal}%
  \BibitemOpen
  \bibfield  {author} {\bibinfo {author} {\bibfnamefont {B.~B.}\ \bibnamefont
  {Mandelbrot}},\ }\href@noop {} {\emph {\bibinfo {title} {The fractal geometry
  of nature}}},\ Vol.~\bibinfo {volume} {1}\ (\bibinfo  {publisher} {WH freeman
  New York},\ \bibinfo {year} {1982})\BibitemShut {NoStop}%
\bibitem [{\citenamefont {Schroeder}(2009)}]{schroeder2009fractals}%
  \BibitemOpen
  \bibfield  {author} {\bibinfo {author} {\bibfnamefont {M.}~\bibnamefont
  {Schroeder}},\ }\href@noop {} {\emph {\bibinfo {title} {Fractals, chaos,
  power laws: Minutes from an infinite paradise}}}\ (\bibinfo  {publisher}
  {Courier Corporation},\ \bibinfo {year} {2009})\BibitemShut {NoStop}%
\bibitem [{\citenamefont {Falconer}(2004)}]{falconer2004fractal}%
  \BibitemOpen
  \bibfield  {author} {\bibinfo {author} {\bibfnamefont {K.}~\bibnamefont
  {Falconer}},\ }\href@noop {} {\emph {\bibinfo {title} {Fractal geometry:
  mathematical foundations and applications}}}\ (\bibinfo  {publisher} {John
  Wiley \& Sons},\ \bibinfo {year} {2004})\BibitemShut {NoStop}%
\bibitem [{\citenamefont {Nakayama}\ and\ \citenamefont
  {Yakubo}(2003)}]{nakayama2003fractal}%
  \BibitemOpen
  \bibfield  {author} {\bibinfo {author} {\bibfnamefont {T.}~\bibnamefont
  {Nakayama}}\ and\ \bibinfo {author} {\bibfnamefont {K.}~\bibnamefont
  {Yakubo}},\ }\href@noop {} {\emph {\bibinfo {title} {Fractal concepts in
  condensed matter physics}}},\ Vol.\ \bibinfo {volume} {140}\ (\bibinfo
  {publisher} {Springer Science \& Business Media},\ \bibinfo {year}
  {2003})\BibitemShut {NoStop}%
\bibitem [{\citenamefont {Mandelbrot}(1977)}]{Mandelbrot1977fractals}%
  \BibitemOpen
  \bibfield  {author} {\bibinfo {author} {\bibfnamefont {B.}~\bibnamefont
  {Mandelbrot}},\ }\href@noop {} {\emph {\bibinfo {title} {Fractals}}}\
  (\bibinfo  {publisher} {Freeman San Francisco},\ \bibinfo {year}
  {1977})\BibitemShut {NoStop}%
\bibitem [{\citenamefont {Sierpinski}(1915)}]{Sierpinski1915Sur}%
  \BibitemOpen
  \bibfield  {author} {\bibinfo {author} {\bibfnamefont {W.}~\bibnamefont
  {Sierpinski}},\ }\bibfield  {title} {\bibinfo {title} {Sur une courbe dont
  tout point est un point de ramification},\ }\href
  {https://cir.nii.ac.jp/crid/1573950399016422528} {\bibfield  {journal}
  {\bibinfo  {journal} {C. R. Acad. Sci.}\ }\textbf {\bibinfo {volume} {160}},\
  \bibinfo {pages} {302} (\bibinfo {year} {1915})}\BibitemShut {NoStop}%
\bibitem [{\citenamefont {Vicsek}(1983)}]{Vicsek1983Fractal}%
  \BibitemOpen
  \bibfield  {author} {\bibinfo {author} {\bibfnamefont {T.}~\bibnamefont
  {Vicsek}},\ }\bibfield  {title} {\bibinfo {title} {Fractal models for
  diffusion controlled aggregation},\ }\href
  {https://doi.org/10.1088/0305-4470/16/17/003} {\bibfield  {journal} {\bibinfo
   {journal} {Journal of Physics A: Mathematical and General}\ }\textbf
  {\bibinfo {volume} {16}},\ \bibinfo {pages} {L647} (\bibinfo {year}
  {1983})}\BibitemShut {NoStop}%
\bibitem [{\citenamefont {You}\ \emph {et~al.}(1993)\citenamefont {You},
  \citenamefont {Lam}, \citenamefont {Nori},\ and\ \citenamefont
  {Sander}}]{You1993Exact}%
  \BibitemOpen
  \bibfield  {author} {\bibinfo {author} {\bibfnamefont {J.~Q.}\ \bibnamefont
  {You}}, \bibinfo {author} {\bibfnamefont {C.-H.}\ \bibnamefont {Lam}},
  \bibinfo {author} {\bibfnamefont {F.}~\bibnamefont {Nori}},\ and\ \bibinfo
  {author} {\bibfnamefont {L.~M.}\ \bibnamefont {Sander}},\ }\bibfield  {title}
  {\bibinfo {title} {Exact renormalization-group approach to the generating
  function of the vicsek fractal},\ }\href
  {https://doi.org/10.1103/PhysRevE.48.R4183} {\bibfield  {journal} {\bibinfo
  {journal} {Phys. Rev. E}\ }\textbf {\bibinfo {volume} {48}},\ \bibinfo
  {pages} {R4183} (\bibinfo {year} {1993})}\BibitemShut {NoStop}%
\bibitem [{\citenamefont {Von~Koch}(1906)}]{von1906methode}%
  \BibitemOpen
  \bibfield  {author} {\bibinfo {author} {\bibfnamefont {H.}~\bibnamefont
  {Von~Koch}},\ }\bibfield  {title} {\bibinfo {title} {Une m{\'e}thode
  g{\'e}om{\'e}trique {\'e}l{\'e}mentaire pour l’{\'e}tude de certaines
  questions de la th{\'e}orie des courbes planes},\ }\href@noop {} {\bibfield
  {journal} {\bibinfo  {journal} {Acta mathematica}\ }\textbf {\bibinfo
  {volume} {30}},\ \bibinfo {pages} {145} (\bibinfo {year} {1906})}\BibitemShut
  {NoStop}%
\bibitem [{\citenamefont {Even}\ \emph {et~al.}(1999)\citenamefont {Even},
  \citenamefont {Russ}, \citenamefont {Repain}, \citenamefont {Pieranski},\
  and\ \citenamefont {Sapoval}}]{Even1999Localizations}%
  \BibitemOpen
  \bibfield  {author} {\bibinfo {author} {\bibfnamefont {C.}~\bibnamefont
  {Even}}, \bibinfo {author} {\bibfnamefont {S.}~\bibnamefont {Russ}}, \bibinfo
  {author} {\bibfnamefont {V.}~\bibnamefont {Repain}}, \bibinfo {author}
  {\bibfnamefont {P.}~\bibnamefont {Pieranski}},\ and\ \bibinfo {author}
  {\bibfnamefont {B.}~\bibnamefont {Sapoval}},\ }\bibfield  {title} {\bibinfo
  {title} {Localizations in fractal drums: An experimental study},\ }\href
  {https://doi.org/10.1103/PhysRevLett.83.726} {\bibfield  {journal} {\bibinfo
  {journal} {Phys. Rev. Lett.}\ }\textbf {\bibinfo {volume} {83}},\ \bibinfo
  {pages} {726} (\bibinfo {year} {1999})}\BibitemShut {NoStop}%
\bibitem [{\citenamefont {Wong}\ and\ \citenamefont
  {Cao}(1992)}]{Wong1992Correlation}%
  \BibitemOpen
  \bibfield  {author} {\bibinfo {author} {\bibfnamefont {P.-z.}\ \bibnamefont
  {Wong}}\ and\ \bibinfo {author} {\bibfnamefont {Q.-z.}\ \bibnamefont {Cao}},\
  }\bibfield  {title} {\bibinfo {title} {Correlation function and structure
  factor for a mass fractal bounded by a surface fractal},\ }\href
  {https://doi.org/10.1103/PhysRevB.45.7627} {\bibfield  {journal} {\bibinfo
  {journal} {Phys. Rev. B}\ }\textbf {\bibinfo {volume} {45}},\ \bibinfo
  {pages} {7627} (\bibinfo {year} {1992})}\BibitemShut {NoStop}%
\bibitem [{\citenamefont {Sapoval}\ and\ \citenamefont
  {Gobron}(1993)}]{Sapoval1993Vibrations}%
  \BibitemOpen
  \bibfield  {author} {\bibinfo {author} {\bibfnamefont {B.}~\bibnamefont
  {Sapoval}}\ and\ \bibinfo {author} {\bibfnamefont {T.}~\bibnamefont
  {Gobron}},\ }\bibfield  {title} {\bibinfo {title} {Vibrations of strongly
  irregular or fractal resonators},\ }\href
  {https://doi.org/10.1103/PhysRevE.47.3013} {\bibfield  {journal} {\bibinfo
  {journal} {Phys. Rev. E}\ }\textbf {\bibinfo {volume} {47}},\ \bibinfo
  {pages} {3013} (\bibinfo {year} {1993})}\BibitemShut {NoStop}%
\bibitem [{\citenamefont {Kopelman}\ \emph {et~al.}(1997)\citenamefont
  {Kopelman}, \citenamefont {Shortreed}, \citenamefont {Shi}, \citenamefont
  {Tan}, \citenamefont {Xu}, \citenamefont {Moore}, \citenamefont {Bar-Haim},\
  and\ \citenamefont {Klafter}}]{Kopelman1997Spectroscopic}%
  \BibitemOpen
  \bibfield  {author} {\bibinfo {author} {\bibfnamefont {R.}~\bibnamefont
  {Kopelman}}, \bibinfo {author} {\bibfnamefont {M.}~\bibnamefont {Shortreed}},
  \bibinfo {author} {\bibfnamefont {Z.-Y.}\ \bibnamefont {Shi}}, \bibinfo
  {author} {\bibfnamefont {W.}~\bibnamefont {Tan}}, \bibinfo {author}
  {\bibfnamefont {Z.}~\bibnamefont {Xu}}, \bibinfo {author} {\bibfnamefont
  {J.~S.}\ \bibnamefont {Moore}}, \bibinfo {author} {\bibfnamefont
  {A.}~\bibnamefont {Bar-Haim}},\ and\ \bibinfo {author} {\bibfnamefont
  {J.}~\bibnamefont {Klafter}},\ }\bibfield  {title} {\bibinfo {title}
  {Spectroscopic evidence for excitonic localization in fractal antenna
  supermolecules},\ }\href {https://doi.org/10.1103/PhysRevLett.78.1239}
  {\bibfield  {journal} {\bibinfo  {journal} {Phys. Rev. Lett.}\ }\textbf
  {\bibinfo {volume} {78}},\ \bibinfo {pages} {1239} (\bibinfo {year}
  {1997})}\BibitemShut {NoStop}%
\bibitem [{\citenamefont {Metz}\ and\ \citenamefont
  {Neri}(2021)}]{Metz2021Localization}%
  \BibitemOpen
  \bibfield  {author} {\bibinfo {author} {\bibfnamefont {F.~L.}\ \bibnamefont
  {Metz}}\ and\ \bibinfo {author} {\bibfnamefont {I.}~\bibnamefont {Neri}},\
  }\bibfield  {title} {\bibinfo {title} {Localization and universality of
  eigenvectors in directed random graphs},\ }\href
  {https://doi.org/10.1103/PhysRevLett.126.040604} {\bibfield  {journal}
  {\bibinfo  {journal} {Phys. Rev. Lett.}\ }\textbf {\bibinfo {volume} {126}},\
  \bibinfo {pages} {040604} (\bibinfo {year} {2021})}\BibitemShut {NoStop}%
\bibitem [{\citenamefont {Matsuki}\ \emph {et~al.}(2021)\citenamefont
  {Matsuki}, \citenamefont {Ikeda},\ and\ \citenamefont
  {Koshino}}]{Matsuki2021Fractal}%
  \BibitemOpen
  \bibfield  {author} {\bibinfo {author} {\bibfnamefont {Y.}~\bibnamefont
  {Matsuki}}, \bibinfo {author} {\bibfnamefont {K.}~\bibnamefont {Ikeda}},\
  and\ \bibinfo {author} {\bibfnamefont {M.}~\bibnamefont {Koshino}},\
  }\bibfield  {title} {\bibinfo {title} {Fractal defect states in the
  hofstadter butterfly},\ }\href {https://doi.org/10.1103/PhysRevB.104.035305}
  {\bibfield  {journal} {\bibinfo  {journal} {Phys. Rev. B}\ }\textbf {\bibinfo
  {volume} {104}},\ \bibinfo {pages} {035305} (\bibinfo {year}
  {2021})}\BibitemShut {NoStop}%
\bibitem [{\citenamefont {You}\ and\ \citenamefont
  {Yang}(1990)}]{You1990Dynamical}%
  \BibitemOpen
  \bibfield  {author} {\bibinfo {author} {\bibfnamefont {J.~Q.}\ \bibnamefont
  {You}}\ and\ \bibinfo {author} {\bibfnamefont {Q.~B.}\ \bibnamefont {Yang}},\
  }\bibfield  {title} {\bibinfo {title} {Dynamical maps and cantor-like spectra
  for a class of one-dimensional quasiperiodic lattices},\ }\href
  {https://doi.org/10.1088/0953-8984/2/8/015} {\bibfield  {journal} {\bibinfo
  {journal} {Journal of Physics: Condensed Matter}\ }\textbf {\bibinfo {volume}
  {2}},\ \bibinfo {pages} {2093} (\bibinfo {year} {1990})}\BibitemShut
  {NoStop}%
\bibitem [{\citenamefont {Pedersen}(2020)}]{Pedersen2020Graphene}%
  \BibitemOpen
  \bibfield  {author} {\bibinfo {author} {\bibfnamefont {T.~G.}\ \bibnamefont
  {Pedersen}},\ }\bibfield  {title} {\bibinfo {title} {Graphene fractals:
  Energy gap and spin polarization},\ }\href
  {https://doi.org/10.1103/PhysRevB.101.235427} {\bibfield  {journal} {\bibinfo
   {journal} {Phys. Rev. B}\ }\textbf {\bibinfo {volume} {101}},\ \bibinfo
  {pages} {235427} (\bibinfo {year} {2020})}\BibitemShut {NoStop}%
\bibitem [{\citenamefont {W\'ojcik}\ \emph {et~al.}(2000)\citenamefont
  {W\'ojcik}, \citenamefont {Bia\l{}ynicki-Birula},\ and\ \citenamefont
  {\ifmmode~\dot{Z}\else \.{Z}\fi{}yczkowski}}]{Wojcik2000Time}%
  \BibitemOpen
  \bibfield  {author} {\bibinfo {author} {\bibfnamefont {D.}~\bibnamefont
  {W\'ojcik}}, \bibinfo {author} {\bibfnamefont {I.}~\bibnamefont
  {Bia\l{}ynicki-Birula}},\ and\ \bibinfo {author} {\bibfnamefont
  {K.}~\bibnamefont {\ifmmode~\dot{Z}\else \.{Z}\fi{}yczkowski}},\ }\bibfield
  {title} {\bibinfo {title} {Time evolution of quantum fractals},\ }\href
  {https://doi.org/10.1103/PhysRevLett.85.5022} {\bibfield  {journal} {\bibinfo
   {journal} {Phys. Rev. Lett.}\ }\textbf {\bibinfo {volume} {85}},\ \bibinfo
  {pages} {5022} (\bibinfo {year} {2000})}\BibitemShut {NoStop}%
\bibitem [{\citenamefont {Rammal}(1983)}]{Rammal1983Nature}%
  \BibitemOpen
  \bibfield  {author} {\bibinfo {author} {\bibfnamefont {R.}~\bibnamefont
  {Rammal}},\ }\bibfield  {title} {\bibinfo {title} {Nature of eigenstates on
  fractal structures},\ }\href {https://doi.org/10.1103/PhysRevB.28.4871}
  {\bibfield  {journal} {\bibinfo  {journal} {Phys. Rev. B}\ }\textbf {\bibinfo
  {volume} {28}},\ \bibinfo {pages} {4871} (\bibinfo {year}
  {1983})}\BibitemShut {NoStop}%
\bibitem [{\citenamefont {Schwalm}\ and\ \citenamefont
  {Schwalm}(1989)}]{Schwalm1989Electronic}%
  \BibitemOpen
  \bibfield  {author} {\bibinfo {author} {\bibfnamefont {W.~A.}\ \bibnamefont
  {Schwalm}}\ and\ \bibinfo {author} {\bibfnamefont {M.~K.}\ \bibnamefont
  {Schwalm}},\ }\bibfield  {title} {\bibinfo {title} {Electronic properties of
  fractal-glass models},\ }\href {https://doi.org/10.1103/PhysRevB.39.12872}
  {\bibfield  {journal} {\bibinfo  {journal} {Phys. Rev. B}\ }\textbf {\bibinfo
  {volume} {39}},\ \bibinfo {pages} {12872} (\bibinfo {year}
  {1989})}\BibitemShut {NoStop}%
\bibitem [{\citenamefont {Schwalm}\ and\ \citenamefont
  {Schwalm}(1993)}]{Schwalm1993Explicit}%
  \BibitemOpen
  \bibfield  {author} {\bibinfo {author} {\bibfnamefont {W.~A.}\ \bibnamefont
  {Schwalm}}\ and\ \bibinfo {author} {\bibfnamefont {M.~K.}\ \bibnamefont
  {Schwalm}},\ }\bibfield  {title} {\bibinfo {title} {Explicit orbits for
  renormalization maps for green functions on fractal lattices},\ }\href
  {https://doi.org/10.1103/PhysRevB.47.7847} {\bibfield  {journal} {\bibinfo
  {journal} {Phys. Rev. B}\ }\textbf {\bibinfo {volume} {47}},\ \bibinfo
  {pages} {7847} (\bibinfo {year} {1993})}\BibitemShut {NoStop}%
\bibitem [{\citenamefont {Penrose}(1974)}]{Penrose1874The}%
  \BibitemOpen
  \bibfield  {author} {\bibinfo {author} {\bibfnamefont {R.}~\bibnamefont
  {Penrose}},\ }\bibfield  {title} {\bibinfo {title} {The role of aesthetics in
  pure and applied mathematical research},\ }\href
  {https://cir.nii.ac.jp/crid/1572543024090822912} {\bibfield  {journal}
  {\bibinfo  {journal} {Bull.Inst.Math.Appl.}\ }\textbf {\bibinfo {volume}
  {10}},\ \bibinfo {pages} {266} (\bibinfo {year} {1974})}\BibitemShut
  {NoStop}%
\bibitem [{\citenamefont {Gardner}(1977)}]{Martin1977math}%
  \BibitemOpen
  \bibfield  {author} {\bibinfo {author} {\bibfnamefont {M.}~\bibnamefont
  {Gardner}},\ }\bibfield  {title} {\bibinfo {title} {Mathematical games},\
  }\href {http://www.jstor.org/stable/24954008} {\bibfield  {journal} {\bibinfo
   {journal} {Scientific American}\ }\textbf {\bibinfo {volume} {237}},\
  \bibinfo {pages} {120} (\bibinfo {year} {1977})}\BibitemShut {NoStop}%
\bibitem [{\citenamefont {Shechtman}\ \emph {et~al.}(1984)\citenamefont
  {Shechtman}, \citenamefont {Blech}, \citenamefont {Gratias},\ and\
  \citenamefont {Cahn}}]{Shechtman1984Metallic}%
  \BibitemOpen
  \bibfield  {author} {\bibinfo {author} {\bibfnamefont {D.}~\bibnamefont
  {Shechtman}}, \bibinfo {author} {\bibfnamefont {I.}~\bibnamefont {Blech}},
  \bibinfo {author} {\bibfnamefont {D.}~\bibnamefont {Gratias}},\ and\ \bibinfo
  {author} {\bibfnamefont {J.~W.}\ \bibnamefont {Cahn}},\ }\bibfield  {title}
  {\bibinfo {title} {Metallic phase with long-range orientational order and no
  translational symmetry},\ }\href
  {https://doi.org/10.1103/PhysRevLett.53.1951} {\bibfield  {journal} {\bibinfo
   {journal} {Phys. Rev. Lett.}\ }\textbf {\bibinfo {volume} {53}},\ \bibinfo
  {pages} {1951} (\bibinfo {year} {1984})}\BibitemShut {NoStop}%
\bibitem [{\citenamefont {Levine}\ and\ \citenamefont
  {Steinhardt}(1984)}]{Levine1984Quasi}%
  \BibitemOpen
  \bibfield  {author} {\bibinfo {author} {\bibfnamefont {D.}~\bibnamefont
  {Levine}}\ and\ \bibinfo {author} {\bibfnamefont {P.~J.}\ \bibnamefont
  {Steinhardt}},\ }\bibfield  {title} {\bibinfo {title} {Quasicrystals: A new
  class of ordered structures},\ }\href
  {https://doi.org/10.1103/PhysRevLett.53.2477} {\bibfield  {journal} {\bibinfo
   {journal} {Phys. Rev. Lett.}\ }\textbf {\bibinfo {volume} {53}},\ \bibinfo
  {pages} {2477} (\bibinfo {year} {1984})}\BibitemShut {NoStop}%
\bibitem [{\citenamefont {Newkome}\ \emph {et~al.}(2006)\citenamefont
  {Newkome}, \citenamefont {Wang}, \citenamefont {Moorefield}, \citenamefont
  {Cho}, \citenamefont {Mohapatra}, \citenamefont {Li}, \citenamefont {Hwang},
  \citenamefont {Lukoyanova}, \citenamefont {Echegoyen}, \citenamefont
  {Palagallo}, \citenamefont {Iancu},\ and\ \citenamefont
  {Hla}}]{George2006Nanoassembly}%
  \BibitemOpen
  \bibfield  {author} {\bibinfo {author} {\bibfnamefont {G.~R.}\ \bibnamefont
  {Newkome}}, \bibinfo {author} {\bibfnamefont {P.}~\bibnamefont {Wang}},
  \bibinfo {author} {\bibfnamefont {C.~N.}\ \bibnamefont {Moorefield}},
  \bibinfo {author} {\bibfnamefont {T.~J.}\ \bibnamefont {Cho}}, \bibinfo
  {author} {\bibfnamefont {P.~P.}\ \bibnamefont {Mohapatra}}, \bibinfo {author}
  {\bibfnamefont {S.}~\bibnamefont {Li}}, \bibinfo {author} {\bibfnamefont
  {S.-H.}\ \bibnamefont {Hwang}}, \bibinfo {author} {\bibfnamefont
  {O.}~\bibnamefont {Lukoyanova}}, \bibinfo {author} {\bibfnamefont
  {L.}~\bibnamefont {Echegoyen}}, \bibinfo {author} {\bibfnamefont {J.~A.}\
  \bibnamefont {Palagallo}}, \bibinfo {author} {\bibfnamefont {V.}~\bibnamefont
  {Iancu}},\ and\ \bibinfo {author} {\bibfnamefont {S.-W.}\ \bibnamefont
  {Hla}},\ }\bibfield  {title} {\bibinfo {title} {Nanoassembly of a fractal
  polymer: A molecular "sierpinski hexagonal gasket"},\ }\href
  {https://doi.org/10.1126/science.1125894} {\bibfield  {journal} {\bibinfo
  {journal} {Science}\ }\textbf {\bibinfo {volume} {312}},\ \bibinfo {pages}
  {1782} (\bibinfo {year} {2006})}\BibitemShut {NoStop}%
\bibitem [{\citenamefont {Fan}\ \emph {et~al.}(2014)\citenamefont {Fan},
  \citenamefont {Yeo}, \citenamefont {Su}, \citenamefont {Hattori},
  \citenamefont {Lee}, \citenamefont {Jung}, \citenamefont {Zhang},
  \citenamefont {Liu}, \citenamefont {Cheng}, \citenamefont {Falgout},
  \citenamefont {Bajema}, \citenamefont {Coleman}, \citenamefont {Gregoire},
  \citenamefont {Larsen}, \citenamefont {Huang},\ and\ \citenamefont
  {Rogers}}]{Fan2014Fractal}%
  \BibitemOpen
  \bibfield  {author} {\bibinfo {author} {\bibfnamefont {J.~A.}\ \bibnamefont
  {Fan}}, \bibinfo {author} {\bibfnamefont {W.-H.}\ \bibnamefont {Yeo}},
  \bibinfo {author} {\bibfnamefont {Y.}~\bibnamefont {Su}}, \bibinfo {author}
  {\bibfnamefont {Y.}~\bibnamefont {Hattori}}, \bibinfo {author} {\bibfnamefont
  {W.}~\bibnamefont {Lee}}, \bibinfo {author} {\bibfnamefont {S.-Y.}\
  \bibnamefont {Jung}}, \bibinfo {author} {\bibfnamefont {Y.}~\bibnamefont
  {Zhang}}, \bibinfo {author} {\bibfnamefont {Z.}~\bibnamefont {Liu}}, \bibinfo
  {author} {\bibfnamefont {H.}~\bibnamefont {Cheng}}, \bibinfo {author}
  {\bibfnamefont {L.}~\bibnamefont {Falgout}}, \bibinfo {author} {\bibfnamefont
  {M.}~\bibnamefont {Bajema}}, \bibinfo {author} {\bibfnamefont
  {T.}~\bibnamefont {Coleman}}, \bibinfo {author} {\bibfnamefont
  {D.}~\bibnamefont {Gregoire}}, \bibinfo {author} {\bibfnamefont {R.~J.}\
  \bibnamefont {Larsen}}, \bibinfo {author} {\bibfnamefont {Y.}~\bibnamefont
  {Huang}},\ and\ \bibinfo {author} {\bibfnamefont {J.~A.}\ \bibnamefont
  {Rogers}},\ }\bibfield  {title} {\bibinfo {title} {Fractal design concepts
  for stretchable electronics},\ }\bibfield  {journal} {\bibinfo  {journal}
  {Nature Communications}\ }\textbf {\bibinfo {volume} {5}},\ \href
  {https://doi.org/10.1038/ncomms4266} {10.1038/ncomms4266} (\bibinfo {year}
  {2014})\BibitemShut {NoStop}%
\bibitem [{\citenamefont {Hernando}\ \emph
  {et~al.}(2015{\natexlab{a}})\citenamefont {Hernando}, \citenamefont {Sulc},\
  and\ \citenamefont {Vanicek}}]{Hernando2015Mesoscale}%
  \BibitemOpen
  \bibfield  {author} {\bibinfo {author} {\bibfnamefont {A.}~\bibnamefont
  {Hernando}}, \bibinfo {author} {\bibfnamefont {M.}~\bibnamefont {Sulc}},\
  and\ \bibinfo {author} {\bibfnamefont {J.}~\bibnamefont {Vanicek}},\ }\href
  {https://doi.org/10.48550/ARXIV.1503.07741} {\bibinfo {title} {Spectral
  properties of electrons in fractal nanowires}} (\bibinfo {year}
  {2015}{\natexlab{a}})\BibitemShut {NoStop}%
\bibitem [{\citenamefont {Shang}\ \emph {et~al.}(2015)\citenamefont {Shang},
  \citenamefont {Wang}, \citenamefont {Chen}, \citenamefont {Dai},
  \citenamefont {Zhou}, \citenamefont {Kuttner}, \citenamefont {Hilt},
  \citenamefont {Shao}, \citenamefont {Gottfried},\ and\ \citenamefont
  {Wu}}]{Shang2015Assembling}%
  \BibitemOpen
  \bibfield  {author} {\bibinfo {author} {\bibfnamefont {J.}~\bibnamefont
  {Shang}}, \bibinfo {author} {\bibfnamefont {Y.}~\bibnamefont {Wang}},
  \bibinfo {author} {\bibfnamefont {M.}~\bibnamefont {Chen}}, \bibinfo {author}
  {\bibfnamefont {J.}~\bibnamefont {Dai}}, \bibinfo {author} {\bibfnamefont
  {X.}~\bibnamefont {Zhou}}, \bibinfo {author} {\bibfnamefont {J.}~\bibnamefont
  {Kuttner}}, \bibinfo {author} {\bibfnamefont {G.}~\bibnamefont {Hilt}},
  \bibinfo {author} {\bibfnamefont {X.}~\bibnamefont {Shao}}, \bibinfo {author}
  {\bibfnamefont {J.~M.}\ \bibnamefont {Gottfried}},\ and\ \bibinfo {author}
  {\bibfnamefont {K.}~\bibnamefont {Wu}},\ }\bibfield  {title} {\bibinfo
  {title} {Assembling molecular sierpi{\'{n}}ski triangle fractals},\ }\href
  {https://doi.org/10.1038/nchem.2211} {\bibfield  {journal} {\bibinfo
  {journal} {Nature Chemistry}\ }\textbf {\bibinfo {volume} {7}},\ \bibinfo
  {pages} {389} (\bibinfo {year} {2015})}\BibitemShut {NoStop}%
\bibitem [{\citenamefont {Zhang}\ \emph {et~al.}(2016)\citenamefont {Zhang},
  \citenamefont {Li}, \citenamefont {Liu}, \citenamefont {Gu}, \citenamefont
  {Li}, \citenamefont {Tang}, \citenamefont {Peng}, \citenamefont {Hou},\ and\
  \citenamefont {Wang}}]{Zhang2016Robust}%
  \BibitemOpen
  \bibfield  {author} {\bibinfo {author} {\bibfnamefont {X.}~\bibnamefont
  {Zhang}}, \bibinfo {author} {\bibfnamefont {N.}~\bibnamefont {Li}}, \bibinfo
  {author} {\bibfnamefont {L.}~\bibnamefont {Liu}}, \bibinfo {author}
  {\bibfnamefont {G.}~\bibnamefont {Gu}}, \bibinfo {author} {\bibfnamefont
  {C.}~\bibnamefont {Li}}, \bibinfo {author} {\bibfnamefont {H.}~\bibnamefont
  {Tang}}, \bibinfo {author} {\bibfnamefont {L.}~\bibnamefont {Peng}}, \bibinfo
  {author} {\bibfnamefont {S.}~\bibnamefont {Hou}},\ and\ \bibinfo {author}
  {\bibfnamefont {Y.}~\bibnamefont {Wang}},\ }\bibfield  {title} {\bibinfo
  {title} {Robust sierpi{\'{n}}ski triangle fractals on symmetry-mismatched
  ag(100)},\ }\href {https://doi.org/10.1039/c6cc04879j} {\bibfield  {journal}
  {\bibinfo  {journal} {Chemical Communications}\ }\textbf {\bibinfo {volume}
  {52}},\ \bibinfo {pages} {10578} (\bibinfo {year} {2016})}\BibitemShut
  {NoStop}%
\bibitem [{\citenamefont {Kempkes}\ \emph {et~al.}(2018)\citenamefont
  {Kempkes}, \citenamefont {Slot}, \citenamefont {Freeney}, \citenamefont
  {Zevenhuizen}, \citenamefont {Vanmaekelbergh}, \citenamefont {Swart},\ and\
  \citenamefont {Smith}}]{Kempkes2018Design}%
  \BibitemOpen
  \bibfield  {author} {\bibinfo {author} {\bibfnamefont {S.~N.}\ \bibnamefont
  {Kempkes}}, \bibinfo {author} {\bibfnamefont {M.~R.}\ \bibnamefont {Slot}},
  \bibinfo {author} {\bibfnamefont {S.~E.}\ \bibnamefont {Freeney}}, \bibinfo
  {author} {\bibfnamefont {S.~J.~M.}\ \bibnamefont {Zevenhuizen}}, \bibinfo
  {author} {\bibfnamefont {D.}~\bibnamefont {Vanmaekelbergh}}, \bibinfo
  {author} {\bibfnamefont {I.}~\bibnamefont {Swart}},\ and\ \bibinfo {author}
  {\bibfnamefont {C.~M.}\ \bibnamefont {Smith}},\ }\bibfield  {title} {\bibinfo
  {title} {Design and characterization of electrons in a fractal geometry},\
  }\href {https://doi.org/10.1038/s41567-018-0328-0} {\bibfield  {journal}
  {\bibinfo  {journal} {Nature Physics}\ }\textbf {\bibinfo {volume} {15}},\
  \bibinfo {pages} {127} (\bibinfo {year} {2018})}\BibitemShut {NoStop}%
\bibitem [{\citenamefont {Liu}\ \emph {et~al.}(2021)\citenamefont {Liu},
  \citenamefont {Zhou}, \citenamefont {Wang}, \citenamefont {Yin},
  \citenamefont {Li}, \citenamefont {Huang}, \citenamefont {Guan},
  \citenamefont {Li}, \citenamefont {Wang}, \citenamefont {Zheng},
  \citenamefont {Liu}, \citenamefont {Han}, \citenamefont {Evans},
  \citenamefont {Liu},\ and\ \citenamefont {Jia}}]{Liu2021Sierpi}%
  \BibitemOpen
  \bibfield  {author} {\bibinfo {author} {\bibfnamefont {C.}~\bibnamefont
  {Liu}}, \bibinfo {author} {\bibfnamefont {Y.}~\bibnamefont {Zhou}}, \bibinfo
  {author} {\bibfnamefont {G.}~\bibnamefont {Wang}}, \bibinfo {author}
  {\bibfnamefont {Y.}~\bibnamefont {Yin}}, \bibinfo {author} {\bibfnamefont
  {C.}~\bibnamefont {Li}}, \bibinfo {author} {\bibfnamefont {H.}~\bibnamefont
  {Huang}}, \bibinfo {author} {\bibfnamefont {D.}~\bibnamefont {Guan}},
  \bibinfo {author} {\bibfnamefont {Y.}~\bibnamefont {Li}}, \bibinfo {author}
  {\bibfnamefont {S.}~\bibnamefont {Wang}}, \bibinfo {author} {\bibfnamefont
  {H.}~\bibnamefont {Zheng}}, \bibinfo {author} {\bibfnamefont
  {C.}~\bibnamefont {Liu}}, \bibinfo {author} {\bibfnamefont {Y.}~\bibnamefont
  {Han}}, \bibinfo {author} {\bibfnamefont {J.~W.}\ \bibnamefont {Evans}},
  \bibinfo {author} {\bibfnamefont {F.}~\bibnamefont {Liu}},\ and\ \bibinfo
  {author} {\bibfnamefont {J.}~\bibnamefont {Jia}},\ }\bibfield  {title}
  {\bibinfo {title} {Sierpi\ifmmode \acute{n}\else \'{n}\fi{}ski structure and
  electronic topology in bi thin films on insb(111)b surfaces},\ }\href
  {https://doi.org/10.1103/PhysRevLett.126.176102} {\bibfield  {journal}
  {\bibinfo  {journal} {Phys. Rev. Lett.}\ }\textbf {\bibinfo {volume} {126}},\
  \bibinfo {pages} {176102} (\bibinfo {year} {2021})}\BibitemShut {NoStop}%
\bibitem [{\citenamefont {Gefen}\ \emph {et~al.}(1980)\citenamefont {Gefen},
  \citenamefont {Mandelbrot},\ and\ \citenamefont
  {Aharony}}]{Gefen1980Critical}%
  \BibitemOpen
  \bibfield  {author} {\bibinfo {author} {\bibfnamefont {Y.}~\bibnamefont
  {Gefen}}, \bibinfo {author} {\bibfnamefont {B.~B.}\ \bibnamefont
  {Mandelbrot}},\ and\ \bibinfo {author} {\bibfnamefont {A.}~\bibnamefont
  {Aharony}},\ }\bibfield  {title} {\bibinfo {title} {Critical phenomena on
  fractal lattices},\ }\href {https://doi.org/10.1103/PhysRevLett.45.855}
  {\bibfield  {journal} {\bibinfo  {journal} {Phys. Rev. Lett.}\ }\textbf
  {\bibinfo {volume} {45}},\ \bibinfo {pages} {855} (\bibinfo {year}
  {1980})}\BibitemShut {NoStop}%
\bibitem [{\citenamefont {Ohta}\ and\ \citenamefont
  {Honjo}(1988)}]{Ohta1988Growth}%
  \BibitemOpen
  \bibfield  {author} {\bibinfo {author} {\bibfnamefont {S.}~\bibnamefont
  {Ohta}}\ and\ \bibinfo {author} {\bibfnamefont {H.}~\bibnamefont {Honjo}},\
  }\bibfield  {title} {\bibinfo {title} {Growth probability distribution in
  irregular fractal-like crystal growth of ammonium chloride},\ }\href
  {https://doi.org/10.1103/PhysRevLett.60.611} {\bibfield  {journal} {\bibinfo
  {journal} {Phys. Rev. Lett.}\ }\textbf {\bibinfo {volume} {60}},\ \bibinfo
  {pages} {611} (\bibinfo {year} {1988})}\BibitemShut {NoStop}%
\bibitem [{\citenamefont {Bo-Ming}\ and\ \citenamefont
  {Kai-Lun}(1988)}]{Yu1988Numerical}%
  \BibitemOpen
  \bibfield  {author} {\bibinfo {author} {\bibfnamefont {Y.}~\bibnamefont
  {Bo-Ming}}\ and\ \bibinfo {author} {\bibfnamefont {Y.}~\bibnamefont
  {Kai-Lun}},\ }\bibfield  {title} {\bibinfo {title} {Numerical evidence of the
  critical percolation probability $p_c =1$ for site problems on sierpinski
  gaskets},\ }\href {https://doi.org/10.1088/0305-4470/21/15/015} {\bibfield
  {journal} {\bibinfo  {journal} {Journal of Physics A: Mathematical and
  General}\ }\textbf {\bibinfo {volume} {21}},\ \bibinfo {pages} {3269}
  (\bibinfo {year} {1988})}\BibitemShut {NoStop}%
\bibitem [{\citenamefont {Suzuki}(1983)}]{Suzuki1983Phase}%
  \BibitemOpen
  \bibfield  {author} {\bibinfo {author} {\bibfnamefont {M.}~\bibnamefont
  {Suzuki}},\ }\bibfield  {title} {\bibinfo {title} {{Phase Transition and
  Fractals}},\ }\href {https://doi.org/10.1143/PTP.69.65} {\bibfield  {journal}
  {\bibinfo  {journal} {Progress of Theoretical Physics}\ }\textbf {\bibinfo
  {volume} {69}},\ \bibinfo {pages} {65} (\bibinfo {year} {1983})}\BibitemShut
  {NoStop}%
\bibitem [{\citenamefont {Gefen}\ \emph {et~al.}(1983)\citenamefont {Gefen},
  \citenamefont {Meir}, \citenamefont {Mandelbrot},\ and\ \citenamefont
  {Aharony}}]{Gefen1983Geometric}%
  \BibitemOpen
  \bibfield  {author} {\bibinfo {author} {\bibfnamefont {Y.}~\bibnamefont
  {Gefen}}, \bibinfo {author} {\bibfnamefont {Y.}~\bibnamefont {Meir}},
  \bibinfo {author} {\bibfnamefont {B.~B.}\ \bibnamefont {Mandelbrot}},\ and\
  \bibinfo {author} {\bibfnamefont {A.}~\bibnamefont {Aharony}},\ }\bibfield
  {title} {\bibinfo {title} {Geometric implementation of hypercubic lattices
  with noninteger dimensionality by use of low lacunarity fractal lattices},\
  }\href {https://doi.org/10.1103/PhysRevLett.50.145} {\bibfield  {journal}
  {\bibinfo  {journal} {Phys. Rev. Lett.}\ }\textbf {\bibinfo {volume} {50}},\
  \bibinfo {pages} {145} (\bibinfo {year} {1983})}\BibitemShut {NoStop}%
\bibitem [{\citenamefont {Lin}\ and\ \citenamefont
  {Yang}(1986)}]{Lin1986suggested}%
  \BibitemOpen
  \bibfield  {author} {\bibinfo {author} {\bibfnamefont {B.}~\bibnamefont
  {Lin}}\ and\ \bibinfo {author} {\bibfnamefont {Z.~R.}\ \bibnamefont {Yang}},\
  }\bibfield  {title} {\bibinfo {title} {A suggested lacunarity expression for
  sierpinski carpets},\ }\href {https://doi.org/10.1088/0305-4470/19/2/005}
  {\bibfield  {journal} {\bibinfo  {journal} {Journal of Physics A:
  Mathematical and General}\ }\textbf {\bibinfo {volume} {19}},\ \bibinfo
  {pages} {L49} (\bibinfo {year} {1986})}\BibitemShut {NoStop}%
\bibitem [{\citenamefont {Taguchi}(1987)}]{Taguchi1987Comment}%
  \BibitemOpen
  \bibfield  {author} {\bibinfo {author} {\bibfnamefont {Y.}~\bibnamefont
  {Taguchi}},\ }\bibfield  {title} {\bibinfo {title} {Lacunarity and
  universality},\ }\href {https://doi.org/10.1088/0305-4470/20/18/058}
  {\bibfield  {journal} {\bibinfo  {journal} {Journal of Physics A:
  Mathematical and General}\ }\textbf {\bibinfo {volume} {20}},\ \bibinfo
  {pages} {6611} (\bibinfo {year} {1987})}\BibitemShut {NoStop}%
\bibitem [{\citenamefont {Lin}(1987)}]{Lin1987Classi}%
  \BibitemOpen
  \bibfield  {author} {\bibinfo {author} {\bibfnamefont {B.}~\bibnamefont
  {Lin}},\ }\bibfield  {title} {\bibinfo {title} {Classification and universal
  properties of sierpinski carpets},\ }\href
  {https://doi.org/10.1088/0305-4470/20/3/009} {\bibfield  {journal} {\bibinfo
  {journal} {Journal of Physics A: Mathematical and General}\ }\textbf
  {\bibinfo {volume} {20}},\ \bibinfo {pages} {L163} (\bibinfo {year}
  {1987})}\BibitemShut {NoStop}%
\bibitem [{\citenamefont {Wu}(1988)}]{Wu1988Comment}%
  \BibitemOpen
  \bibfield  {author} {\bibinfo {author} {\bibfnamefont {Y.-K.}\ \bibnamefont
  {Wu}},\ }\bibfield  {title} {\bibinfo {title} {Lacunarity and universality},\
  }\href {https://doi.org/10.1088/0305-4470/21/22/027} {\bibfield  {journal}
  {\bibinfo  {journal} {Journal of Physics A: Mathematical and General}\
  }\textbf {\bibinfo {volume} {21}},\ \bibinfo {pages} {4251} (\bibinfo {year}
  {1988})}\BibitemShut {NoStop}%
\bibitem [{\citenamefont {van Veen}\ \emph {et~al.}(2016)\citenamefont {van
  Veen}, \citenamefont {Yuan}, \citenamefont {Katsnelson}, \citenamefont
  {Polini},\ and\ \citenamefont {Tomadin}}]{Edo2016Transport}%
  \BibitemOpen
  \bibfield  {author} {\bibinfo {author} {\bibfnamefont {E.}~\bibnamefont {van
  Veen}}, \bibinfo {author} {\bibfnamefont {S.}~\bibnamefont {Yuan}}, \bibinfo
  {author} {\bibfnamefont {M.~I.}\ \bibnamefont {Katsnelson}}, \bibinfo
  {author} {\bibfnamefont {M.}~\bibnamefont {Polini}},\ and\ \bibinfo {author}
  {\bibfnamefont {A.}~\bibnamefont {Tomadin}},\ }\bibfield  {title} {\bibinfo
  {title} {Quantum transport in sierpinski carpets},\ }\href
  {https://doi.org/10.1103/PhysRevB.93.115428} {\bibfield  {journal} {\bibinfo
  {journal} {Phys. Rev. B}\ }\textbf {\bibinfo {volume} {93}},\ \bibinfo
  {pages} {115428} (\bibinfo {year} {2016})}\BibitemShut {NoStop}%
\bibitem [{\citenamefont {Xu}\ \emph {et~al.}(2021)\citenamefont {Xu},
  \citenamefont {Wang}, \citenamefont {Chen}, \citenamefont {Smith},\ and\
  \citenamefont {Jin}}]{Xu2021Quantum}%
  \BibitemOpen
  \bibfield  {author} {\bibinfo {author} {\bibfnamefont {X.-Y.}\ \bibnamefont
  {Xu}}, \bibinfo {author} {\bibfnamefont {X.-W.}\ \bibnamefont {Wang}},
  \bibinfo {author} {\bibfnamefont {D.-Y.}\ \bibnamefont {Chen}}, \bibinfo
  {author} {\bibfnamefont {C.~M.}\ \bibnamefont {Smith}},\ and\ \bibinfo
  {author} {\bibfnamefont {X.-M.}\ \bibnamefont {Jin}},\ }\bibfield  {title}
  {\bibinfo {title} {Quantum transport in fractal networks},\ }\href
  {https://doi.org/10.1038/s41566-021-00845-4} {\bibfield  {journal} {\bibinfo
  {journal} {Nature Photonics}\ }\textbf {\bibinfo {volume} {15}},\ \bibinfo
  {pages} {703} (\bibinfo {year} {2021})}\BibitemShut {NoStop}%
\bibitem [{\citenamefont {Iliasov}\ \emph
  {et~al.}(2020{\natexlab{a}})\citenamefont {Iliasov}, \citenamefont
  {Katsnelson},\ and\ \citenamefont {Yuan}}]{Askar2020Hall}%
  \BibitemOpen
  \bibfield  {author} {\bibinfo {author} {\bibfnamefont {A.~A.}\ \bibnamefont
  {Iliasov}}, \bibinfo {author} {\bibfnamefont {M.~I.}\ \bibnamefont
  {Katsnelson}},\ and\ \bibinfo {author} {\bibfnamefont {S.}~\bibnamefont
  {Yuan}},\ }\bibfield  {title} {\bibinfo {title} {Hall conductivity of a
  sierpi\ifmmode \acute{n}\else \'{n}\fi{}ski carpet},\ }\href
  {https://doi.org/10.1103/PhysRevB.101.045413} {\bibfield  {journal} {\bibinfo
   {journal} {Phys. Rev. B}\ }\textbf {\bibinfo {volume} {101}},\ \bibinfo
  {pages} {045413} (\bibinfo {year} {2020}{\natexlab{a}})}\BibitemShut
  {NoStop}%
\bibitem [{\citenamefont {van Veen}\ \emph {et~al.}(2017)\citenamefont {van
  Veen}, \citenamefont {Tomadin}, \citenamefont {Polini}, \citenamefont
  {Katsnelson},\ and\ \citenamefont {Yuan}}]{Edo2017Optical}%
  \BibitemOpen
  \bibfield  {author} {\bibinfo {author} {\bibfnamefont {E.}~\bibnamefont {van
  Veen}}, \bibinfo {author} {\bibfnamefont {A.}~\bibnamefont {Tomadin}},
  \bibinfo {author} {\bibfnamefont {M.}~\bibnamefont {Polini}}, \bibinfo
  {author} {\bibfnamefont {M.~I.}\ \bibnamefont {Katsnelson}},\ and\ \bibinfo
  {author} {\bibfnamefont {S.}~\bibnamefont {Yuan}},\ }\bibfield  {title}
  {\bibinfo {title} {Optical conductivity of a quantum electron gas in a
  sierpinski carpet},\ }\href {https://doi.org/10.1103/PhysRevB.96.235438}
  {\bibfield  {journal} {\bibinfo  {journal} {Phys. Rev. B}\ }\textbf {\bibinfo
  {volume} {96}},\ \bibinfo {pages} {235438} (\bibinfo {year}
  {2017})}\BibitemShut {NoStop}%
\bibitem [{\citenamefont {Westerhout}\ \emph {et~al.}(2018)\citenamefont
  {Westerhout}, \citenamefont {van Veen}, \citenamefont {Katsnelson},\ and\
  \citenamefont {Yuan}}]{Tom2018Plasmon}%
  \BibitemOpen
  \bibfield  {author} {\bibinfo {author} {\bibfnamefont {T.}~\bibnamefont
  {Westerhout}}, \bibinfo {author} {\bibfnamefont {E.}~\bibnamefont {van
  Veen}}, \bibinfo {author} {\bibfnamefont {M.~I.}\ \bibnamefont
  {Katsnelson}},\ and\ \bibinfo {author} {\bibfnamefont {S.}~\bibnamefont
  {Yuan}},\ }\bibfield  {title} {\bibinfo {title} {Plasmon confinement in
  fractal quantum systems},\ }\href
  {https://doi.org/10.1103/PhysRevB.97.205434} {\bibfield  {journal} {\bibinfo
  {journal} {Phys. Rev. B}\ }\textbf {\bibinfo {volume} {97}},\ \bibinfo
  {pages} {205434} (\bibinfo {year} {2018})}\BibitemShut {NoStop}%
\bibitem [{\citenamefont {Gefen}\ \emph {et~al.}(1984)\citenamefont {Gefen},
  \citenamefont {Aharony},\ and\ \citenamefont {Mandelbrot}}]{Gefen1984Phase}%
  \BibitemOpen
  \bibfield  {author} {\bibinfo {author} {\bibfnamefont {Y.}~\bibnamefont
  {Gefen}}, \bibinfo {author} {\bibfnamefont {A.}~\bibnamefont {Aharony}},\
  and\ \bibinfo {author} {\bibfnamefont {B.~B.}\ \bibnamefont {Mandelbrot}},\
  }\bibfield  {title} {\bibinfo {title} {Phase transitions on fractals. {III}.
  infinitely ramified lattices},\ }\href
  {https://doi.org/10.1088/0305-4470/17/6/024} {\bibfield  {journal} {\bibinfo
  {journal} {Journal of Physics A: Mathematical and General}\ }\textbf
  {\bibinfo {volume} {17}},\ \bibinfo {pages} {1277} (\bibinfo {year}
  {1984})}\BibitemShut {NoStop}%
\bibitem [{\citenamefont {Akkermans}\ \emph {et~al.}(2010)\citenamefont
  {Akkermans}, \citenamefont {Dunne},\ and\ \citenamefont
  {Teplyaev}}]{Akkermans2010Thermodynamics}%
  \BibitemOpen
  \bibfield  {author} {\bibinfo {author} {\bibfnamefont {E.}~\bibnamefont
  {Akkermans}}, \bibinfo {author} {\bibfnamefont {G.~V.}\ \bibnamefont
  {Dunne}},\ and\ \bibinfo {author} {\bibfnamefont {A.}~\bibnamefont
  {Teplyaev}},\ }\bibfield  {title} {\bibinfo {title} {Thermodynamics of
  photons on fractals},\ }\href
  {https://doi.org/10.1103/PhysRevLett.105.230407} {\bibfield  {journal}
  {\bibinfo  {journal} {Phys. Rev. Lett.}\ }\textbf {\bibinfo {volume} {105}},\
  \bibinfo {pages} {230407} (\bibinfo {year} {2010})}\BibitemShut {NoStop}%
\bibitem [{\citenamefont {Rammal}(1984)}]{Rammal1984Spectrum}%
  \BibitemOpen
  \bibfield  {author} {\bibinfo {author} {\bibfnamefont {R.}~\bibnamefont
  {Rammal}},\ }\bibfield  {title} {\bibinfo {title} {{Spectrum of harmonic
  excitations on fractals}},\ }\href
  {https://doi.org/10.1051/jphys:01984004502019100} {\bibfield  {journal}
  {\bibinfo  {journal} {{Journal de Physique}}\ }\textbf {\bibinfo {volume}
  {45}},\ \bibinfo {pages} {191} (\bibinfo {year} {1984})}\BibitemShut
  {NoStop}%
\bibitem [{\citenamefont {Pal}\ \emph {et~al.}(2013)\citenamefont {Pal},
  \citenamefont {Patra}, \citenamefont {Saha},\ and\ \citenamefont
  {Chakrabarti}}]{Pal2013Engineering}%
  \BibitemOpen
  \bibfield  {author} {\bibinfo {author} {\bibfnamefont {B.}~\bibnamefont
  {Pal}}, \bibinfo {author} {\bibfnamefont {P.}~\bibnamefont {Patra}}, \bibinfo
  {author} {\bibfnamefont {J.~P.}\ \bibnamefont {Saha}},\ and\ \bibinfo
  {author} {\bibfnamefont {A.}~\bibnamefont {Chakrabarti}},\ }\bibfield
  {title} {\bibinfo {title} {Engineering wave localization in a fractal
  waveguide network},\ }\href {https://doi.org/10.1103/PhysRevA.87.023814}
  {\bibfield  {journal} {\bibinfo  {journal} {Phys. Rev. A}\ }\textbf {\bibinfo
  {volume} {87}},\ \bibinfo {pages} {023814} (\bibinfo {year}
  {2013})}\BibitemShut {NoStop}%
\bibitem [{\citenamefont {Xia}\ \emph {et~al.}(2022)\citenamefont {Xia},
  \citenamefont {Huang}, \citenamefont {Wang},\ and\ \citenamefont
  {Li}}]{Xia2022Exact}%
  \BibitemOpen
  \bibfield  {author} {\bibinfo {author} {\bibfnamefont {X.}~\bibnamefont
  {Xia}}, \bibinfo {author} {\bibfnamefont {K.}~\bibnamefont {Huang}}, \bibinfo
  {author} {\bibfnamefont {S.}~\bibnamefont {Wang}},\ and\ \bibinfo {author}
  {\bibfnamefont {X.}~\bibnamefont {Li}},\ }\bibfield  {title} {\bibinfo
  {title} {Exact mobility edges in the non-hermitian
  ${t}_{1}\text{\ensuremath{-}}{t}_{2}$ model: Theory and possible experimental
  realizations},\ }\href {https://doi.org/10.1103/PhysRevB.105.014207}
  {\bibfield  {journal} {\bibinfo  {journal} {Phys. Rev. B}\ }\textbf {\bibinfo
  {volume} {105}},\ \bibinfo {pages} {014207} (\bibinfo {year}
  {2022})}\BibitemShut {NoStop}%
\bibitem [{\citenamefont {Yang}\ \emph {et~al.}(2020)\citenamefont {Yang},
  \citenamefont {Zhou}, \citenamefont {Zhao},\ and\ \citenamefont
  {Yuan}}]{Yang2020Confined}%
  \BibitemOpen
  \bibfield  {author} {\bibinfo {author} {\bibfnamefont {X.}~\bibnamefont
  {Yang}}, \bibinfo {author} {\bibfnamefont {W.}~\bibnamefont {Zhou}}, \bibinfo
  {author} {\bibfnamefont {P.}~\bibnamefont {Zhao}},\ and\ \bibinfo {author}
  {\bibfnamefont {S.}~\bibnamefont {Yuan}},\ }\bibfield  {title} {\bibinfo
  {title} {Confined electrons in effective plane fractals},\ }\href
  {https://doi.org/10.1103/PhysRevB.102.245425} {\bibfield  {journal} {\bibinfo
   {journal} {Phys. Rev. B}\ }\textbf {\bibinfo {volume} {102}},\ \bibinfo
  {pages} {245425} (\bibinfo {year} {2020})}\BibitemShut {NoStop}%
\bibitem [{\citenamefont {Balankin}(2017)}]{Alexander2017The}%
  \BibitemOpen
  \bibfield  {author} {\bibinfo {author} {\bibfnamefont {A.~S.}\ \bibnamefont
  {Balankin}},\ }\bibfield  {title} {\bibinfo {title} {The topological
  hausdorff dimension and transport properties of sierpiński carpets},\ }\href
  {https://doi.org/https://doi.org/10.1016/j.physleta.2017.06.049} {\bibfield
  {journal} {\bibinfo  {journal} {Physics Letters A}\ }\textbf {\bibinfo
  {volume} {381}},\ \bibinfo {pages} {2801} (\bibinfo {year}
  {2017})}\BibitemShut {NoStop}%
\bibitem [{\citenamefont {Feder}(2013)}]{feder2013fractals}%
  \BibitemOpen
  \bibfield  {author} {\bibinfo {author} {\bibfnamefont {J.}~\bibnamefont
  {Feder}},\ }\href@noop {} {\emph {\bibinfo {title} {Fractals}}}\ (\bibinfo
  {publisher} {Springer Science \& Business Media},\ \bibinfo {year}
  {2013})\BibitemShut {NoStop}%
\bibitem [{\citenamefont {Akemann}\ \emph {et~al.}(2015)\citenamefont
  {Akemann}, \citenamefont {Baik},\ and\ \citenamefont
  {Di~Francesco}}]{Akemann2015RMT}%
  \BibitemOpen
  \bibfield  {author} {\bibinfo {author} {\bibfnamefont {G.}~\bibnamefont
  {Akemann}}, \bibinfo {author} {\bibfnamefont {J.}~\bibnamefont {Baik}},\ and\
  \bibinfo {author} {\bibfnamefont {P.}~\bibnamefont {Di~Francesco}},\ }\href
  {https://doi.org/10.1093/oxfordhb/9780198744191.001.0001} {\emph {\bibinfo
  {title} {{The Oxford Handbook of Random Matrix Theory}}}}\ (\bibinfo
  {publisher} {Oxford University Press},\ \bibinfo {year} {2015})\BibitemShut
  {NoStop}%
\bibitem [{\citenamefont {Iliasov}\ \emph
  {et~al.}(2020{\natexlab{b}})\citenamefont {Iliasov}, \citenamefont
  {Katsnelson},\ and\ \citenamefont {Yuan}}]{Iliasov2020Linearized}%
  \BibitemOpen
  \bibfield  {author} {\bibinfo {author} {\bibfnamefont {A.~A.}\ \bibnamefont
  {Iliasov}}, \bibinfo {author} {\bibfnamefont {M.~I.}\ \bibnamefont
  {Katsnelson}},\ and\ \bibinfo {author} {\bibfnamefont {S.}~\bibnamefont
  {Yuan}},\ }\bibfield  {title} {\bibinfo {title} {Linearized spectral
  decimation in fractals},\ }\href
  {https://doi.org/10.1103/PhysRevB.102.075440} {\bibfield  {journal} {\bibinfo
   {journal} {Phys. Rev. B}\ }\textbf {\bibinfo {volume} {102}},\ \bibinfo
  {pages} {075440} (\bibinfo {year} {2020}{\natexlab{b}})}\BibitemShut
  {NoStop}%
\bibitem [{\citenamefont {Iliasov}\ \emph {et~al.}(2019)\citenamefont
  {Iliasov}, \citenamefont {Katsnelson},\ and\ \citenamefont
  {Yuan}}]{Askar2019Power}%
  \BibitemOpen
  \bibfield  {author} {\bibinfo {author} {\bibfnamefont {A.~A.}\ \bibnamefont
  {Iliasov}}, \bibinfo {author} {\bibfnamefont {M.~I.}\ \bibnamefont
  {Katsnelson}},\ and\ \bibinfo {author} {\bibfnamefont {S.}~\bibnamefont
  {Yuan}},\ }\bibfield  {title} {\bibinfo {title} {Power-law energy level
  spacing distributions in fractals},\ }\href
  {https://doi.org/10.1103/PhysRevB.99.075402} {\bibfield  {journal} {\bibinfo
  {journal} {Phys. Rev. B}\ }\textbf {\bibinfo {volume} {99}},\ \bibinfo
  {pages} {075402} (\bibinfo {year} {2019})}\BibitemShut {NoStop}%
\bibitem [{\citenamefont {Dyson}(1962{\natexlab{a}})}]{Dyson1962I}%
  \BibitemOpen
  \bibfield  {author} {\bibinfo {author} {\bibfnamefont {F.~J.}\ \bibnamefont
  {Dyson}},\ }\bibfield  {title} {\bibinfo {title} {Statistical theory of the
  energy levels of complex systems. i},\ }\href@noop {} {\bibfield  {journal}
  {\bibinfo  {journal} {Journal of Mathematical Physics}\ }\textbf {\bibinfo
  {volume} {3}},\ \bibinfo {pages} {140} (\bibinfo {year}
  {1962}{\natexlab{a}})}\BibitemShut {NoStop}%
\bibitem [{\citenamefont {Dyson}(1962{\natexlab{b}})}]{Dyson1962II}%
  \BibitemOpen
  \bibfield  {author} {\bibinfo {author} {\bibfnamefont {F.~J.}\ \bibnamefont
  {Dyson}},\ }\bibfield  {title} {\bibinfo {title} {Statistical theory of the
  energy levels of complex systems. {II}},\ }\href
  {https://doi.org/10.1063/1.1703774} {\bibfield  {journal} {\bibinfo
  {journal} {Journal of Mathematical Physics}\ }\textbf {\bibinfo {volume}
  {3}},\ \bibinfo {pages} {157} (\bibinfo {year}
  {1962}{\natexlab{b}})}\BibitemShut {NoStop}%
\bibitem [{\citenamefont {Dyson}(1962{\natexlab{c}})}]{Dyson1962III}%
  \BibitemOpen
  \bibfield  {author} {\bibinfo {author} {\bibfnamefont {F.~J.}\ \bibnamefont
  {Dyson}},\ }\bibfield  {title} {\bibinfo {title} {Statistical theory of the
  energy levels of complex systems. iii},\ }\href
  {https://doi.org/10.1063/1.1703775} {\bibfield  {journal} {\bibinfo
  {journal} {Journal of Mathematical Physics}\ }\textbf {\bibinfo {volume}
  {3}},\ \bibinfo {pages} {166} (\bibinfo {year}
  {1962}{\natexlab{c}})}\BibitemShut {NoStop}%
\bibitem [{\citenamefont {Yao}\ \emph {et~al.}(2022{\natexlab{a}})\citenamefont
  {Yao}, \citenamefont {Yang}, \citenamefont {A.~Iliasov}, \citenamefont
  {Katsnelson},\ and\ \citenamefont {Yuan}}]{Yao2022Wave}%
  \BibitemOpen
  \bibfield  {author} {\bibinfo {author} {\bibfnamefont {Q.}~\bibnamefont
  {Yao}}, \bibinfo {author} {\bibfnamefont {X.-T.}\ \bibnamefont {Yang}},
  \bibinfo {author} {\bibfnamefont {A.}~\bibnamefont {A.~Iliasov}}, \bibinfo
  {author} {\bibfnamefont {M.~I.}\ \bibnamefont {Katsnelson}},\ and\ \bibinfo
  {author} {\bibfnamefont {S.}~\bibnamefont {Yuan}},\ }\href
  {https://doi.org/update lately} {\bibinfo {title} {Wave functions in critical
  phase: a planar \textit{Sierpi\'{n}ski} fractal lattice (to be submitted)}}
  (\bibinfo {year} {2022}{\natexlab{a}})\BibitemShut {NoStop}%
\bibitem [{\citenamefont {Yao}\ \emph {et~al.}(2022{\natexlab{b}})\citenamefont
  {Yao}, \citenamefont {Katsnelson},\ and\ \citenamefont {Yuan}}]{Yao2022New}%
  \BibitemOpen
  \bibfield  {author} {\bibinfo {author} {\bibfnamefont {Q.}~\bibnamefont
  {Yao}}, \bibinfo {author} {\bibfnamefont {M.~I.}\ \bibnamefont
  {Katsnelson}},\ and\ \bibinfo {author} {\bibfnamefont {S.}~\bibnamefont
  {Yuan}},\ }\href {https://doi.org/update lately} {\bibinfo {title} {New
  localization in scale-invariance fractal system (to be submitted)}} (\bibinfo
  {year} {2022}{\natexlab{b}})\BibitemShut {NoStop}%
\bibitem [{Com()}]{CompuMax}%
  \BibitemOpen
  \href@noop {} {}\bibinfo {note} {Here we can estimate that when $g$ increases
  by a unit, the value $N=Dim(H)$ extends $8$ times for the $self$ pattern of
  $\mathcal{N}=8$, however for the $gene$ pattern, it is larger and depends on
  the special choice of the $generator(n,m)$, up to $n^2-m^2$ times. It brings
  the unaffordable computational overhead at time o($N^3$) and at memory
  o($N^2$)}\BibitemShut {NoStop}%
\bibitem [{Deg()}]{Degeneratenote}%
  \BibitemOpen
  \href@noop {} {}\bibinfo {note} {For a quasidegenerated (or degenerated)
  level cluster, their states tend to be non-orthogonal between each other, and
  the spatial overlapping in fractal lattice causes the correlation between
  these levels. Using the participation ratio or generalized multifractal
  analysis, we further clarify the diversity between these critical
  states.}\BibitemShut {Stop}%
\bibitem [{\citenamefont {Machida}\ and\ \citenamefont
  {Fujita}(1986)}]{Machida1986Quantum}%
  \BibitemOpen
  \bibfield  {author} {\bibinfo {author} {\bibfnamefont {K.}~\bibnamefont
  {Machida}}\ and\ \bibinfo {author} {\bibfnamefont {M.}~\bibnamefont
  {Fujita}},\ }\bibfield  {title} {\bibinfo {title} {Quantum energy spectra and
  one-dimensional quasiperiodic systems},\ }\href
  {https://doi.org/10.1103/PhysRevB.34.7367} {\bibfield  {journal} {\bibinfo
  {journal} {Phys. Rev. B}\ }\textbf {\bibinfo {volume} {34}},\ \bibinfo
  {pages} {7367} (\bibinfo {year} {1986})}\BibitemShut {NoStop}%
\bibitem [{\citenamefont {Bloch}(1929)}]{Bloch1929}%
  \BibitemOpen
  \bibfield  {author} {\bibinfo {author} {\bibfnamefont {F.}~\bibnamefont
  {Bloch}},\ }\bibfield  {title} {\bibinfo {title} {On the quantum mechanics of
  electrons in crystal lattices},\ }\href {https://doi.org/10.1007/bf01339455}
  {\bibfield  {journal} {\bibinfo  {journal} {magazine for physics}\ }\textbf
  {\bibinfo {volume} {52}},\ \bibinfo {pages} {555} (\bibinfo {year}
  {1929})}\BibitemShut {NoStop}%
\bibitem [{\citenamefont {Anderson}(1958)}]{Anderson1958Absence}%
  \BibitemOpen
  \bibfield  {author} {\bibinfo {author} {\bibfnamefont {P.~W.}\ \bibnamefont
  {Anderson}},\ }\bibfield  {title} {\bibinfo {title} {Absence of diffusion in
  certain random lattices},\ }\href {https://doi.org/10.1103/PhysRev.109.1492}
  {\bibfield  {journal} {\bibinfo  {journal} {Phys. Rev.}\ }\textbf {\bibinfo
  {volume} {109}},\ \bibinfo {pages} {1492} (\bibinfo {year}
  {1958})}\BibitemShut {NoStop}%
\bibitem [{\citenamefont {Howland}(1987)}]{HOWLAND1987Perturbation}%
  \BibitemOpen
  \bibfield  {author} {\bibinfo {author} {\bibfnamefont {J.~S.}\ \bibnamefont
  {Howland}},\ }\bibfield  {title} {\bibinfo {title} {Perturbation theory of
  dense point spectra},\ }\href
  {https://doi.org/https://doi.org/10.1016/0022-1236(87)90038-3} {\bibfield
  {journal} {\bibinfo  {journal} {Journal of Functional Analysis}\ }\textbf
  {\bibinfo {volume} {74}},\ \bibinfo {pages} {52} (\bibinfo {year}
  {1987})}\BibitemShut {NoStop}%
\bibitem [{\citenamefont {Hiramoto}\ and\ \citenamefont
  {Kohmoto}(1989)}]{Hiramoto1989New}%
  \BibitemOpen
  \bibfield  {author} {\bibinfo {author} {\bibfnamefont {H.}~\bibnamefont
  {Hiramoto}}\ and\ \bibinfo {author} {\bibfnamefont {M.}~\bibnamefont
  {Kohmoto}},\ }\bibfield  {title} {\bibinfo {title} {New localization in a
  quasiperiodic system},\ }\href {https://doi.org/10.1103/PhysRevLett.62.2714}
  {\bibfield  {journal} {\bibinfo  {journal} {Phys. Rev. Lett.}\ }\textbf
  {\bibinfo {volume} {62}},\ \bibinfo {pages} {2714} (\bibinfo {year}
  {1989})}\BibitemShut {NoStop}%
\bibitem [{\citenamefont {Fujiwara}\ \emph {et~al.}(1989)\citenamefont
  {Fujiwara}, \citenamefont {Kohmoto},\ and\ \citenamefont
  {Tokihiro}}]{Fujiwara1989Multifractal}%
  \BibitemOpen
  \bibfield  {author} {\bibinfo {author} {\bibfnamefont {T.}~\bibnamefont
  {Fujiwara}}, \bibinfo {author} {\bibfnamefont {M.}~\bibnamefont {Kohmoto}},\
  and\ \bibinfo {author} {\bibfnamefont {T.}~\bibnamefont {Tokihiro}},\
  }\bibfield  {title} {\bibinfo {title} {Multifractal wave functions on a
  fibonacci lattice},\ }\href {https://doi.org/10.1103/PhysRevB.40.7413}
  {\bibfield  {journal} {\bibinfo  {journal} {Phys. Rev. B}\ }\textbf {\bibinfo
  {volume} {40}},\ \bibinfo {pages} {7413} (\bibinfo {year}
  {1989})}\BibitemShut {NoStop}%
\bibitem [{\citenamefont {hiramoto}\ and\ \citenamefont
  {Kohmoto}(1992)}]{Hisashi1992}%
  \BibitemOpen
  \bibfield  {author} {\bibinfo {author} {\bibfnamefont {H.}~\bibnamefont
  {hiramoto}}\ and\ \bibinfo {author} {\bibfnamefont {M.}~\bibnamefont
  {Kohmoto}},\ }\bibfield  {title} {\bibinfo {title} {Electronic spectral and
  wavefunction properties of one-dimensional quasiperiodic systems: a scaling
  approach},\ }\href {https://doi.org/10.1142/s0217979292000153} {\bibfield
  {journal} {\bibinfo  {journal} {International Journal of Modern Physics B}\
  }\textbf {\bibinfo {volume} {06}},\ \bibinfo {pages} {281} (\bibinfo {year}
  {1992})}\BibitemShut {NoStop}%
\bibitem [{\citenamefont {Vidal}\ \emph {et~al.}(1998)\citenamefont {Vidal},
  \citenamefont {Mosseri},\ and\ \citenamefont {Dou\ifmmode~\mbox{\c{c}}\else
  \c{c}\fi{}ot}}]{Vidal1998AB}%
  \BibitemOpen
  \bibfield  {author} {\bibinfo {author} {\bibfnamefont {J.}~\bibnamefont
  {Vidal}}, \bibinfo {author} {\bibfnamefont {R.}~\bibnamefont {Mosseri}},\
  and\ \bibinfo {author} {\bibfnamefont {B.}~\bibnamefont
  {Dou\ifmmode~\mbox{\c{c}}\else \c{c}\fi{}ot}},\ }\bibfield  {title} {\bibinfo
  {title} {Aharonov-bohm cages in two-dimensional structures},\ }\href
  {https://doi.org/10.1103/PhysRevLett.81.5888} {\bibfield  {journal} {\bibinfo
   {journal} {Phys. Rev. Lett.}\ }\textbf {\bibinfo {volume} {81}},\ \bibinfo
  {pages} {5888} (\bibinfo {year} {1998})}\BibitemShut {NoStop}%
\bibitem [{\citenamefont {Vidal}\ \emph {et~al.}(2001)\citenamefont {Vidal},
  \citenamefont {Butaud}, \citenamefont {Dou\ifmmode~\mbox{\c{c}}\else
  \c{c}\fi{}ot},\ and\ \citenamefont {Mosseri}}]{Vidal2001Disorder}%
  \BibitemOpen
  \bibfield  {author} {\bibinfo {author} {\bibfnamefont {J.}~\bibnamefont
  {Vidal}}, \bibinfo {author} {\bibfnamefont {P.}~\bibnamefont {Butaud}},
  \bibinfo {author} {\bibfnamefont {B.}~\bibnamefont
  {Dou\ifmmode~\mbox{\c{c}}\else \c{c}\fi{}ot}},\ and\ \bibinfo {author}
  {\bibfnamefont {R.}~\bibnamefont {Mosseri}},\ }\bibfield  {title} {\bibinfo
  {title} {Disorder and interactions in aharonov-bohm cages},\ }\href
  {https://doi.org/10.1103/PhysRevB.64.155306} {\bibfield  {journal} {\bibinfo
  {journal} {Phys. Rev. B}\ }\textbf {\bibinfo {volume} {64}},\ \bibinfo
  {pages} {155306} (\bibinfo {year} {2001})}\BibitemShut {NoStop}%
\bibitem [{\citenamefont {Mukherjee}\ \emph {et~al.}(2018)\citenamefont
  {Mukherjee}, \citenamefont {Di~Liberto}, \citenamefont {\"Ohberg},
  \citenamefont {Thomson},\ and\ \citenamefont
  {Goldman}}]{Mukherjee2018Experimental}%
  \BibitemOpen
  \bibfield  {author} {\bibinfo {author} {\bibfnamefont {S.}~\bibnamefont
  {Mukherjee}}, \bibinfo {author} {\bibfnamefont {M.}~\bibnamefont
  {Di~Liberto}}, \bibinfo {author} {\bibfnamefont {P.}~\bibnamefont
  {\"Ohberg}}, \bibinfo {author} {\bibfnamefont {R.~R.}\ \bibnamefont
  {Thomson}},\ and\ \bibinfo {author} {\bibfnamefont {N.}~\bibnamefont
  {Goldman}},\ }\bibfield  {title} {\bibinfo {title} {Experimental observation
  of aharonov-bohm cages in photonic lattices},\ }\href
  {https://doi.org/10.1103/PhysRevLett.121.075502} {\bibfield  {journal}
  {\bibinfo  {journal} {Phys. Rev. Lett.}\ }\textbf {\bibinfo {volume} {121}},\
  \bibinfo {pages} {075502} (\bibinfo {year} {2018})}\BibitemShut {NoStop}%
\bibitem [{\citenamefont {Lesser}\ and\ \citenamefont
  {Lifshitz}(2022)}]{Lesser2022Emergence}%
  \BibitemOpen
  \bibfield  {author} {\bibinfo {author} {\bibfnamefont {O.}~\bibnamefont
  {Lesser}}\ and\ \bibinfo {author} {\bibfnamefont {R.}~\bibnamefont
  {Lifshitz}},\ }\bibfield  {title} {\bibinfo {title} {Emergence of
  quasiperiodic bloch wave functions in quasicrystals},\ }\href
  {https://doi.org/10.1103/PhysRevResearch.4.013226} {\bibfield  {journal}
  {\bibinfo  {journal} {Phys. Rev. Research}\ }\textbf {\bibinfo {volume}
  {4}},\ \bibinfo {pages} {013226} (\bibinfo {year} {2022})}\BibitemShut
  {NoStop}%
\bibitem [{\citenamefont {Dyson}\ and\ \citenamefont
  {Mehta}(1963)}]{dyson1963random}%
  \BibitemOpen
  \bibfield  {author} {\bibinfo {author} {\bibfnamefont {F.~J.}\ \bibnamefont
  {Dyson}}\ and\ \bibinfo {author} {\bibfnamefont {M.~L.}\ \bibnamefont
  {Mehta}},\ }\bibfield  {title} {\bibinfo {title} {Statistical theory of the
  energy levels of complex systems. {IV}},\ }\href
  {https://doi.org/10.1063/1.1704008} {\bibfield  {journal} {\bibinfo
  {journal} {Journal of Mathematical Physics}\ }\textbf {\bibinfo {volume}
  {4}},\ \bibinfo {pages} {701} (\bibinfo {year} {1963})}\BibitemShut {NoStop}%
\bibitem [{\citenamefont {Guhr}\ \emph {et~al.}(1998)\citenamefont {Guhr},
  \citenamefont {Müller{\textendash}Groeling},\ and\ \citenamefont
  {Weidenmüller}}]{Guhr1998Random}%
  \BibitemOpen
  \bibfield  {author} {\bibinfo {author} {\bibfnamefont {T.}~\bibnamefont
  {Guhr}}, \bibinfo {author} {\bibfnamefont {A.}~\bibnamefont
  {Müller{\textendash}Groeling}},\ and\ \bibinfo {author} {\bibfnamefont
  {H.~A.}\ \bibnamefont {Weidenmüller}},\ }\bibfield  {title} {\bibinfo
  {title} {Random-matrix theories in quantum physics: common concepts},\ }\href
  {https://doi.org/10.1016/s0370-1573(97)00088-4} {\bibfield  {journal}
  {\bibinfo  {journal} {Physics Reports}\ }\textbf {\bibinfo {volume} {299}},\
  \bibinfo {pages} {189} (\bibinfo {year} {1998})}\BibitemShut {NoStop}%
\bibitem [{\citenamefont {Brody}\ \emph {et~al.}(1981)\citenamefont {Brody},
  \citenamefont {Flores}, \citenamefont {French}, \citenamefont {Mello},
  \citenamefont {Pandey},\ and\ \citenamefont {Wong}}]{Brody1981Random}%
  \BibitemOpen
  \bibfield  {author} {\bibinfo {author} {\bibfnamefont {T.~A.}\ \bibnamefont
  {Brody}}, \bibinfo {author} {\bibfnamefont {J.}~\bibnamefont {Flores}},
  \bibinfo {author} {\bibfnamefont {J.~B.}\ \bibnamefont {French}}, \bibinfo
  {author} {\bibfnamefont {P.~A.}\ \bibnamefont {Mello}}, \bibinfo {author}
  {\bibfnamefont {A.}~\bibnamefont {Pandey}},\ and\ \bibinfo {author}
  {\bibfnamefont {S.~S.~M.}\ \bibnamefont {Wong}},\ }\bibfield  {title}
  {\bibinfo {title} {Random-matrix physics: spectrum and strength
  fluctuations},\ }\href {https://doi.org/10.1103/RevModPhys.53.385} {\bibfield
   {journal} {\bibinfo  {journal} {Rev. Mod. Phys.}\ }\textbf {\bibinfo
  {volume} {53}},\ \bibinfo {pages} {385} (\bibinfo {year} {1981})}\BibitemShut
  {NoStop}%
\bibitem [{\citenamefont {Torres-Herrera}\ \emph {et~al.}(2019)\citenamefont
  {Torres-Herrera}, \citenamefont {M\'endez-Berm\'udez},\ and\ \citenamefont
  {Santos}}]{Torres2019Level}%
  \BibitemOpen
  \bibfield  {author} {\bibinfo {author} {\bibfnamefont {E.~J.}\ \bibnamefont
  {Torres-Herrera}}, \bibinfo {author} {\bibfnamefont {J.~A.}\ \bibnamefont
  {M\'endez-Berm\'udez}},\ and\ \bibinfo {author} {\bibfnamefont {L.~F.}\
  \bibnamefont {Santos}},\ }\bibfield  {title} {\bibinfo {title} {Level
  repulsion and dynamics in the finite one-dimensional anderson model},\ }\href
  {https://doi.org/10.1103/PhysRevE.100.022142} {\bibfield  {journal} {\bibinfo
   {journal} {Phys. Rev. E}\ }\textbf {\bibinfo {volume} {100}},\ \bibinfo
  {pages} {022142} (\bibinfo {year} {2019})}\BibitemShut {NoStop}%
\bibitem [{\citenamefont {G\'omez}\ \emph {et~al.}(2002)\citenamefont
  {G\'omez}, \citenamefont {Molina}, \citenamefont {Rela\~no},\ and\
  \citenamefont {Retamosa}}]{Gomez2002Misleading}%
  \BibitemOpen
  \bibfield  {author} {\bibinfo {author} {\bibfnamefont {J.~M.~G.}\
  \bibnamefont {G\'omez}}, \bibinfo {author} {\bibfnamefont {R.~A.}\
  \bibnamefont {Molina}}, \bibinfo {author} {\bibfnamefont {A.}~\bibnamefont
  {Rela\~no}},\ and\ \bibinfo {author} {\bibfnamefont {J.}~\bibnamefont
  {Retamosa}},\ }\bibfield  {title} {\bibinfo {title} {Misleading signatures of
  quantum chaos},\ }\href {https://doi.org/10.1103/PhysRevE.66.036209}
  {\bibfield  {journal} {\bibinfo  {journal} {Phys. Rev. E}\ }\textbf {\bibinfo
  {volume} {66}},\ \bibinfo {pages} {036209} (\bibinfo {year}
  {2002})}\BibitemShut {NoStop}%
\bibitem [{\citenamefont {Zharekeshev}\ and\ \citenamefont
  {Kramer}(1995)}]{Zharekeshev1995Scaling}%
  \BibitemOpen
  \bibfield  {author} {\bibinfo {author} {\bibfnamefont {I.~K.}\ \bibnamefont
  {Zharekeshev}}\ and\ \bibinfo {author} {\bibfnamefont {B.}~\bibnamefont
  {Kramer}},\ }\bibfield  {title} {\bibinfo {title} {Scaling of level
  statistics at the disorder-induced metal-insulator transition},\ }\href
  {https://doi.org/10.1103/PhysRevB.51.17239} {\bibfield  {journal} {\bibinfo
  {journal} {Phys. Rev. B}\ }\textbf {\bibinfo {volume} {51}},\ \bibinfo
  {pages} {17239} (\bibinfo {year} {1995})}\BibitemShut {NoStop}%
\bibitem [{\citenamefont {Oganesyan}\ and\ \citenamefont
  {Huse}(2007)}]{Ogan2007Local}%
  \BibitemOpen
  \bibfield  {author} {\bibinfo {author} {\bibfnamefont {V.}~\bibnamefont
  {Oganesyan}}\ and\ \bibinfo {author} {\bibfnamefont {D.~A.}\ \bibnamefont
  {Huse}},\ }\bibfield  {title} {\bibinfo {title} {Localization of interacting
  fermions at high temperature},\ }\href
  {https://doi.org/10.1103/PhysRevB.75.155111} {\bibfield  {journal} {\bibinfo
  {journal} {Phys. Rev. B}\ }\textbf {\bibinfo {volume} {75}},\ \bibinfo
  {pages} {155111} (\bibinfo {year} {2007})}\BibitemShut {NoStop}%
\bibitem [{\citenamefont {Giraud}\ \emph {et~al.}(2022)\citenamefont {Giraud},
  \citenamefont {Mac\'e}, \citenamefont {Vernier},\ and\ \citenamefont
  {Alet}}]{Olivier2022Probing}%
  \BibitemOpen
  \bibfield  {author} {\bibinfo {author} {\bibfnamefont {O.}~\bibnamefont
  {Giraud}}, \bibinfo {author} {\bibfnamefont {N.}~\bibnamefont {Mac\'e}},
  \bibinfo {author} {\bibfnamefont {E.}~\bibnamefont {Vernier}},\ and\ \bibinfo
  {author} {\bibfnamefont {F.}~\bibnamefont {Alet}},\ }\bibfield  {title}
  {\bibinfo {title} {Probing symmetries of quantum many-body systems through
  gap ratio statistics},\ }\href {https://doi.org/10.1103/PhysRevX.12.011006}
  {\bibfield  {journal} {\bibinfo  {journal} {Phys. Rev. X}\ }\textbf {\bibinfo
  {volume} {12}},\ \bibinfo {pages} {011006} (\bibinfo {year}
  {2022})}\BibitemShut {NoStop}%
\bibitem [{\citenamefont {S\'a}\ \emph {et~al.}(2020)\citenamefont {S\'a},
  \citenamefont {Ribeiro},\ and\ \citenamefont {Prosen}}]{Lucas2020Complex}%
  \BibitemOpen
  \bibfield  {author} {\bibinfo {author} {\bibfnamefont {L.}~\bibnamefont
  {S\'a}}, \bibinfo {author} {\bibfnamefont {P.}~\bibnamefont {Ribeiro}},\ and\
  \bibinfo {author} {\bibfnamefont {T.~c.~v.}\ \bibnamefont {Prosen}},\
  }\bibfield  {title} {\bibinfo {title} {Complex spacing ratios: A signature of
  dissipative quantum chaos},\ }\href
  {https://doi.org/10.1103/PhysRevX.10.021019} {\bibfield  {journal} {\bibinfo
  {journal} {Phys. Rev. X}\ }\textbf {\bibinfo {volume} {10}},\ \bibinfo
  {pages} {021019} (\bibinfo {year} {2020})}\BibitemShut {NoStop}%
\bibitem [{\citenamefont {Brody}(1973)}]{Brody1973A}%
  \BibitemOpen
  \bibfield  {author} {\bibinfo {author} {\bibfnamefont {T.~A.}\ \bibnamefont
  {Brody}},\ }\bibfield  {title} {\bibinfo {title} {A statistical measure for
  the repulsion of energy levels},\ }\href {https://doi.org/10.1007/bf02727859}
  {\bibfield  {journal} {\bibinfo  {journal} {Lettere Al Nuovo Cimento Series
  2}\ }\textbf {\bibinfo {volume} {7}},\ \bibinfo {pages} {482} (\bibinfo
  {year} {1973})}\BibitemShut {NoStop}%
\bibitem [{\citenamefont {Casati}\ \emph {et~al.}(1991)\citenamefont {Casati},
  \citenamefont {Izrailev},\ and\ \citenamefont
  {Molinari}}]{Casati1991Scaling}%
  \BibitemOpen
  \bibfield  {author} {\bibinfo {author} {\bibfnamefont {G.}~\bibnamefont
  {Casati}}, \bibinfo {author} {\bibfnamefont {F.}~\bibnamefont {Izrailev}},\
  and\ \bibinfo {author} {\bibfnamefont {L.}~\bibnamefont {Molinari}},\
  }\bibfield  {title} {\bibinfo {title} {Scaling properties of the eigenvalue
  spacing distribution for band random matrices},\ }\href
  {https://doi.org/10.1088/0305-4470/24/20/011} {\bibfield  {journal} {\bibinfo
   {journal} {Journal of Physics A: Mathematical and General}\ }\textbf
  {\bibinfo {volume} {24}},\ \bibinfo {pages} {4755} (\bibinfo {year}
  {1991})}\BibitemShut {NoStop}%
\bibitem [{\citenamefont {Aronov}\ \emph {et~al.}(1994)\citenamefont {Aronov},
  \citenamefont {Kravtsov},\ and\ \citenamefont {Lerner}}]{Aronov1994pis}%
  \BibitemOpen
  \bibfield  {author} {\bibinfo {author} {\bibfnamefont {A.}~\bibnamefont
  {Aronov}}, \bibinfo {author} {\bibfnamefont {V.}~\bibnamefont {Kravtsov}},\
  and\ \bibinfo {author} {\bibfnamefont {I.}~\bibnamefont {Lerner}},\
  }\bibfield  {title} {\bibinfo {title} {Pis' ma zh. eksp. teor. fiz. 59: 39
  (1994)},\ }\href@noop {} {\bibfield  {journal} {\bibinfo  {journal} {JETP
  Lett}\ }\textbf {\bibinfo {volume} {59}},\ \bibinfo {pages} {40} (\bibinfo
  {year} {1994})}\BibitemShut {NoStop}%
\bibitem [{\citenamefont {Varga}\ \emph {et~al.}(1995)\citenamefont {Varga},
  \citenamefont {Hofstetter}, \citenamefont {Schreiber},\ and\ \citenamefont
  {Pipek}}]{Varga1995Shape}%
  \BibitemOpen
  \bibfield  {author} {\bibinfo {author} {\bibfnamefont {I.}~\bibnamefont
  {Varga}}, \bibinfo {author} {\bibfnamefont {E.}~\bibnamefont {Hofstetter}},
  \bibinfo {author} {\bibfnamefont {M.}~\bibnamefont {Schreiber}},\ and\
  \bibinfo {author} {\bibfnamefont {J.}~\bibnamefont {Pipek}},\ }\bibfield
  {title} {\bibinfo {title} {Shape analysis of the level-spacing distribution
  around the metal-insulator transition in the three-dimensional anderson
  model},\ }\href {https://doi.org/10.1103/PhysRevB.52.7783} {\bibfield
  {journal} {\bibinfo  {journal} {Phys. Rev. B}\ }\textbf {\bibinfo {volume}
  {52}},\ \bibinfo {pages} {7783} (\bibinfo {year} {1995})}\BibitemShut
  {NoStop}%
\bibitem [{\citenamefont {Atas}\ \emph {et~al.}(2013)\citenamefont {Atas},
  \citenamefont {Bogomolny}, \citenamefont {Giraud},\ and\ \citenamefont
  {Roux}}]{Atas2013Distribution}%
  \BibitemOpen
  \bibfield  {author} {\bibinfo {author} {\bibfnamefont {Y.~Y.}\ \bibnamefont
  {Atas}}, \bibinfo {author} {\bibfnamefont {E.}~\bibnamefont {Bogomolny}},
  \bibinfo {author} {\bibfnamefont {O.}~\bibnamefont {Giraud}},\ and\ \bibinfo
  {author} {\bibfnamefont {G.}~\bibnamefont {Roux}},\ }\bibfield  {title}
  {\bibinfo {title} {Distribution of the ratio of consecutive level spacings in
  random matrix ensembles},\ }\href
  {https://doi.org/10.1103/PhysRevLett.110.084101} {\bibfield  {journal}
  {\bibinfo  {journal} {Phys. Rev. Lett.}\ }\textbf {\bibinfo {volume} {110}},\
  \bibinfo {pages} {084101} (\bibinfo {year} {2013})}\BibitemShut {NoStop}%
\bibitem [{\citenamefont {De~Marco}\ \emph {et~al.}(2022)\citenamefont
  {De~Marco}, \citenamefont {Tolle}, \citenamefont {Halati}, \citenamefont
  {Sheikhan}, \citenamefont {L\"auchli},\ and\ \citenamefont
  {Kollath}}]{De2022Level}%
  \BibitemOpen
  \bibfield  {author} {\bibinfo {author} {\bibfnamefont {J.}~\bibnamefont
  {De~Marco}}, \bibinfo {author} {\bibfnamefont {L.}~\bibnamefont {Tolle}},
  \bibinfo {author} {\bibfnamefont {C.-M.}\ \bibnamefont {Halati}}, \bibinfo
  {author} {\bibfnamefont {A.}~\bibnamefont {Sheikhan}}, \bibinfo {author}
  {\bibfnamefont {A.~M.}\ \bibnamefont {L\"auchli}},\ and\ \bibinfo {author}
  {\bibfnamefont {C.}~\bibnamefont {Kollath}},\ }\bibfield  {title} {\bibinfo
  {title} {Level statistics of the one-dimensional ionic hubbard model},\
  }\href {https://doi.org/10.1103/PhysRevResearch.4.033119} {\bibfield
  {journal} {\bibinfo  {journal} {Phys. Rev. Research}\ }\textbf {\bibinfo
  {volume} {4}},\ \bibinfo {pages} {033119} (\bibinfo {year}
  {2022})}\BibitemShut {NoStop}%
\bibitem [{\citenamefont {Hernando}\ \emph
  {et~al.}(2015{\natexlab{b}})\citenamefont {Hernando}, \citenamefont {Sulc},\
  and\ \citenamefont {Vanicek}}]{Hernando2015Spectral}%
  \BibitemOpen
  \bibfield  {author} {\bibinfo {author} {\bibfnamefont {A.}~\bibnamefont
  {Hernando}}, \bibinfo {author} {\bibfnamefont {M.}~\bibnamefont {Sulc}},\
  and\ \bibinfo {author} {\bibfnamefont {J.}~\bibnamefont {Vanicek}},\ }\href
  {https://doi.org/10.48550/ARXIV.1503.07741} {\bibinfo {title} {Spectral
  properties of electrons in fractal nanowires}} (\bibinfo {year}
  {2015}{\natexlab{b}})\BibitemShut {NoStop}%
\bibitem [{\citenamefont {Domany}\ \emph {et~al.}(1983)\citenamefont {Domany},
  \citenamefont {Alexander}, \citenamefont {Bensimon},\ and\ \citenamefont
  {Kadanoff}}]{Domany1983Solutions}%
  \BibitemOpen
  \bibfield  {author} {\bibinfo {author} {\bibfnamefont {E.}~\bibnamefont
  {Domany}}, \bibinfo {author} {\bibfnamefont {S.}~\bibnamefont {Alexander}},
  \bibinfo {author} {\bibfnamefont {D.}~\bibnamefont {Bensimon}},\ and\
  \bibinfo {author} {\bibfnamefont {L.~P.}\ \bibnamefont {Kadanoff}},\
  }\bibfield  {title} {\bibinfo {title} {Solutions to the schr\"odinger
  equation on some fractal lattices},\ }\href
  {https://doi.org/10.1103/PhysRevB.28.3110} {\bibfield  {journal} {\bibinfo
  {journal} {Phys. Rev. B}\ }\textbf {\bibinfo {volume} {28}},\ \bibinfo
  {pages} {3110} (\bibinfo {year} {1983})}\BibitemShut {NoStop}%
\bibitem [{\citenamefont {You}\ \emph {et~al.}(1992)\citenamefont {You},
  \citenamefont {Yan}, \citenamefont {Zhong},\ and\ \citenamefont
  {Yan}}]{You1992Local}%
  \BibitemOpen
  \bibfield  {author} {\bibinfo {author} {\bibfnamefont {J.~Q.}\ \bibnamefont
  {You}}, \bibinfo {author} {\bibfnamefont {J.~R.}\ \bibnamefont {Yan}},
  \bibinfo {author} {\bibfnamefont {J.~X.}\ \bibnamefont {Zhong}},\ and\
  \bibinfo {author} {\bibfnamefont {X.~H.}\ \bibnamefont {Yan}},\ }\bibfield
  {title} {\bibinfo {title} {Local electronic properties of two-dimensional
  penrose tilings: A renormalization-group approach},\ }\href
  {https://doi.org/10.1103/PhysRevB.45.7690} {\bibfield  {journal} {\bibinfo
  {journal} {Phys. Rev. B}\ }\textbf {\bibinfo {volume} {45}},\ \bibinfo
  {pages} {7690} (\bibinfo {year} {1992})}\BibitemShut {NoStop}%
\bibitem [{\citenamefont {Harper}(1955{\natexlab{a}})}]{Harper1955Single}%
  \BibitemOpen
  \bibfield  {author} {\bibinfo {author} {\bibfnamefont {P.~G.}\ \bibnamefont
  {Harper}},\ }\bibfield  {title} {\bibinfo {title} {Single band motion of
  conduction electrons in a uniform magnetic field},\ }\href
  {https://doi.org/10.1088/0370-1298/68/10/304} {\bibfield  {journal} {\bibinfo
   {journal} {Proceedings of the Physical Society. Section A}\ }\textbf
  {\bibinfo {volume} {68}},\ \bibinfo {pages} {874} (\bibinfo {year}
  {1955}{\natexlab{a}})}\BibitemShut {NoStop}%
\bibitem [{\citenamefont {Aubry}\ and\ \citenamefont
  {Andr{\'e}}(1980)}]{Aubry1980analyticity}%
  \BibitemOpen
  \bibfield  {author} {\bibinfo {author} {\bibfnamefont {S.}~\bibnamefont
  {Aubry}}\ and\ \bibinfo {author} {\bibfnamefont {G.}~\bibnamefont
  {Andr{\'e}}},\ }\bibfield  {title} {\bibinfo {title} {Analyticity breaking
  and anderson localization in incommensurate lattices},\ }\href@noop {}
  {\bibfield  {journal} {\bibinfo  {journal} {Ann. Israel Phys. Soc}\ }\textbf
  {\bibinfo {volume} {3}},\ \bibinfo {pages} {18} (\bibinfo {year}
  {1980})}\BibitemShut {NoStop}%
\bibitem [{\citenamefont {Harper}(1955{\natexlab{b}})}]{Harper1955General}%
  \BibitemOpen
  \bibfield  {author} {\bibinfo {author} {\bibfnamefont {P.~G.}\ \bibnamefont
  {Harper}},\ }\bibfield  {title} {\bibinfo {title} {The general motion of
  conduction electrons in a uniform magnetic field, with application to the
  diamagnetism of metals},\ }\href
  {https://doi.org/10.1088/0370-1298/68/10/305} {\bibfield  {journal} {\bibinfo
   {journal} {Proceedings of the Physical Society. Section A}\ }\textbf
  {\bibinfo {volume} {68}},\ \bibinfo {pages} {879} (\bibinfo {year}
  {1955}{\natexlab{b}})}\BibitemShut {NoStop}%
\bibitem [{\citenamefont {Salazar}\ \emph {et~al.}(2003)\citenamefont
  {Salazar}, \citenamefont {Wang}, \citenamefont {Gelover-Santiago},
  \citenamefont {Zentella-Dehesa}, \citenamefont {Naumis},\ and\ \citenamefont
  {Talamantes}}]{Salazar2003Phonon}%
  \BibitemOpen
  \bibfield  {author} {\bibinfo {author} {\bibfnamefont {F.}~\bibnamefont
  {Salazar}}, \bibinfo {author} {\bibfnamefont {C.}~\bibnamefont {Wang}},
  \bibinfo {author} {\bibfnamefont {A.}~\bibnamefont {Gelover-Santiago}},
  \bibinfo {author} {\bibfnamefont {A.}~\bibnamefont {Zentella-Dehesa}},
  \bibinfo {author} {\bibfnamefont {G.}~\bibnamefont {Naumis}},\ and\ \bibinfo
  {author} {\bibfnamefont {J.}~\bibnamefont {Talamantes}},\ }\bibfield  {title}
  {\bibinfo {title} {Phonon localization in quasiperiodic systems},\ }\href
  {https://doi.org/https://doi.org/10.1016/j.jnoncrysol.2003.08.034} {\bibfield
   {journal} {\bibinfo  {journal} {Journal of Non-Crystalline Solids}\ }\textbf
  {\bibinfo {volume} {329}},\ \bibinfo {pages} {167} (\bibinfo {year}
  {2003})},\ \bibinfo {note} {7th Int. Workshop on Non-Crystalline
  Solids}\BibitemShut {NoStop}%
\bibitem [{\citenamefont {Lin}\ \emph {et~al.}(2016)\citenamefont {Lin},
  \citenamefont {Saad},\ and\ \citenamefont {Yang}}]{Lin2016approximating}%
  \BibitemOpen
  \bibfield  {author} {\bibinfo {author} {\bibfnamefont {L.}~\bibnamefont
  {Lin}}, \bibinfo {author} {\bibfnamefont {Y.}~\bibnamefont {Saad}},\ and\
  \bibinfo {author} {\bibfnamefont {C.}~\bibnamefont {Yang}},\ }\bibfield
  {title} {\bibinfo {title} {Approximating spectral densities of large
  matrices},\ }\href {http://www.jstor.org/stable/24778840} {\bibfield
  {journal} {\bibinfo  {journal} {SIAM Review}\ }\textbf {\bibinfo {volume}
  {58}},\ \bibinfo {pages} {34} (\bibinfo {year} {2016})}\BibitemShut {NoStop}%
\end{thebibliography}

\end{document}